\documentclass{aa}  

\usepackage{soul}
\usepackage{graphicx}
\usepackage{subfigure}
\usepackage{xcolor}
\usepackage{txfonts}
\usepackage{natbib}
\usepackage{siunitx}
\usepackage{lineno}
\usepackage[english]{babel}
\usepackage{hyperref}
\usepackage[normalem]{ulem}

\newcommand{\rlight}{r_{\rm L}}

\def\cor#1{\textcolor{black}{#1}}
\def\cortwo#1{\textcolor{black}{#1}}

\DeclareRobustCommand{\rchi}{{\mathpalette\irchi\relax}}
\newcommand{\irchi}[2]{\raisebox{\depth}{$#1\,\chi$}} 

\DeclareSIUnit\year{yr}

\usepackage{amsmath, amssymb}
\usepackage{listings}
\usepackage{comment}

\begin{document}

\title{The spin-orbit alignment hypothesis in millisecond pulsars}


\author{A. Lorange \inst{1} 
    \and J. P\'etri \inst{1}
    \and M. Sautron \inst{1}
    \and V. Vigon \inst{2,3}
}

\institute{
Universit\'e de Strasbourg, CNRS, Observatoire astronomique de Strasbourg, UMR 7550, F-67000 Strasbourg, France.
\and Université de Strasbourg, CNRS, IRMA,  UMR 7501, F-67000 Strasbourg, France.
\and INRIA, équipe MACARON: Apprentissage automatique pour des méthodes numériques optimisées. F-67000 Strasbourg, France.    \\
	\email{alexandra.lorange@astro.unistra.fr}         
}

  \date{Received ; accepted }

 
 \abstract
   {Millisecond pulsars (MSPs) are spun up during their accretion phase in a binary system. The exchange of angular momentum between the accretion disk and the star tends to align the spin and orbital angular momenta on a very short time scale compared to the accretion stage.}
   {In this work, we study a subset of $\gamma$-ray MSPs in binaries for which the orbital inclination \cor{angle}~$i$ has been accurately constrained thanks to the Shapiro delay measurements. Our goal is to constrain the observer viewing angle~$\zeta$ and to check whether it agrees with the orbital inclination \cor{angle}~$i$, \cor{in other words if} $\zeta \approx i$.}
   {We use a Bayesian inference technique to fit the MSP $\gamma$-ray light curves based on the third $\gamma$-ray pulsar catalogue (3PC). The emission model relies on the striped wind model deduced from force-free neutron star magnetosphere simulations.}
   {We found good agreement between \cor{the} two angles $i$ and $\zeta$ for a significant fraction of our sample, \cor{about four fifth}, confirming the spin-orbit alignment scenario during the accretion stage. However about one fifth of our sample deviates significantly from this alignment. The reasons are manifold: either the $\gamma$-ray fit is not reliable or some precession and external torque avoid an almost perfect alignment.}
   {}


   \maketitle
%

\nolinenumbers
\section{Introduction}

Millisecond pulsars (MSPs) represent a special class of neutron stars rotating at periods around several milliseconds, close to their rotational break up limit. They are spun up during the accretion phase in the binary system they are formed. The accretion efficiency depends on the timescale of this phase, on the type of binary and on nature of the companion star. MSPs spend their life either as isolated pulsars or as accreting X-ray pulsars \citep{lorimer_binary_2008}. These two classes are linked by the recently discovered transitional pulsars. Isolated MSPs are mostly detected as radio-loud $\gamma$-ray pulsars as reported in the third pulsar catalogue \cor{(3PC)} by \cite{smith_third_2023} whereas accreting X-ray pulsars show pulsations in the X-ray band without any counterpart in radio or $\gamma$-rays. In some binary systems with favourable orbital inclination~$i$, that is the angle between the orbital angular momentum of the binary and the line-of-sight, the Shapiro \cor{delay} is detected and used to accurately estimate this \cor{inclination angle}~$i$. Moreover, if the observer line-of-sight angle~$\zeta$, that is the angle between the rotation axis and the line-of-sight, is assumed to coincide with $i$, then powerful tools exist to determine the magnetic obliquity~$\rchi$ from multi-wavelength pulse profile modelling, the obliquity being the angle between the rotation and magnetic axis. 

This idea \cor{of constraining the pulsar geometry from $\gamma$-ray light curves fitting} has been successfully applied to isolated $\gamma$-ray pulsars, like in \cite{benli_constraining_2021} for MSPs and in \cite{petri_young_2021} for young pulsars for which good radio polarisation data even better constrain the geometric angles $\zeta$ and $\rchi$. If we set $\zeta=i$, as expected in a perfect spin-orbit alignment scenario, then the constrain on the obliquity~$\rchi$ will be tighter. 
%
%
Pulse profile modelling (PPM) in X-ray binary pulsars also furnishes an interesting mean to check for stellar spin and orbital angular momentum alignment in those systems, as demonstrated by \cite{laycock_pulse-profile_2025}. Thermal X-ray PPM has also proven to be an efficient tool to deduce the hot spot geometry onto the neutron star surface and to extract the line-of-sight inclination~$\zeta$ \cor{by using the Neutron star Interior Composition ExploreR (NICER) data}, see for instance \cite{riley_nicer_2019, salmi_radius_2024, choudhury_nicer_2024} \cor{for one approach and \cite{miller_psr_2019, miller_radius_2021, dittmann_more_2024} for a different pipeline}.

During the accretion phase in binary systems, the concomitant stellar spin axis and magnetic obliquity evolution involves a complex interaction between the accretion disk and the neutron star magnetosphere. \cor{This interaction includes} electromagnetic torques and \cor{the} precession of a possibly deformed star, deviating from a perfect sphere. The impact on the magnetic inclination angle has been described by \cite{biryukov_magnetic_2021}. A recent review on the spin evolution of neutron star taking into account the different regimes of accretor, propeller and ejector is given by \cite{abolmasov_spin_2024}. Also \cite{yang_magnetic_2023} investigated the evolution of the magnetic obliquity $\rchi$ based on the model of \cite{biryukov_magnetic_2021}. 


From an observational point of view, \cite{guillemot_non-detection_2014} used constrains from binary evolution scenarios to fit MSP $\gamma$-rays light curves. However, only the outer gap and slot gap models were considered. They claimed that the non detection in $\gamma$-rays is due to the unfavourable geometry. However, a flux too low for the Fermi/LAT \cor{(Large Area Telescope)} sensitivity could also explain the lack of detection. Therefore, it is desirable to extend their work by increasing the sample of pulsars seen in $\gamma$-ray and for which Shapiro delays are measured. As the striped wind is now \cor{considered to be} the paradigm for high-energy emission \cor{since the seminal work of \cite{kirk_pulsed_2002} that was later followed up and improved by many authors \citep{petri_unified_2011, contopoulos_pulsar_2010, bai_modeling_2010, kalapotharakos_gamma-ray_2014, cerutti_modelling_2016, kalapotharakos_three-dimensional_2018, philippov_ab-initio_2018, cao_modeling_2019, cao_pulsar_2022, kalapotharakos_gamma-ray_2023, cerutti_synthetic_2025}}, their results must be updated \cor{to take} this emission model \cor{into account}. Among all possible target pulsars, spiders are an interesting subclass of MSPs for which \cite{blanchard_census_2025} published a complete census in our galaxy as observed with the Nançay Radio Telescope (NRT). Many of them show eclipses \citep{clark_neutron_2023, polzin_study_2020} and are also detected in $\gamma$-rays \citep{hui_high_2019}, \cor{a mandatory requirement for our work}. 

In this paper we check the hypothesis of spin-orbit alignment in some MSPs for which the inclination angle~$i$ is accurately constrained by the Shapiro delay or at least constrained with reasonable accuracy. The line-of-sight inclination will be estimated by the $\gamma$-ray light curve fittings.
Sec.~\ref{sec:Model} briefly recalls the underlying model used for the magnetosphere and the associated emission models for radio and $\gamma$-rays. Our Bayesian inference approach and light curve interpolation by discrete Fourier transform is outlined in Sec.~\ref{sec:Method}. Detailed results for individual pulsars are discussed in Sec.~\ref{sec:Results}. \cortwo{The impact of a possible non-radial $\gamma$-ray emission is discussed in Sec.~\ref{sec:discussion-non-radial-prop}}. Possible deviations from perfect alignment are discussed in Sec.~\ref{sec:Discussion}. Finally conclusions are drawn in Sec.~\ref{sec:Conclusion}.

%
%

\section{Emission model and samples\label{sec:Model}}

The multi-wavelength emission model relies on force-free simulations of a dipolar magnetosphere based on \cite{petri_pulsar_2012} and updated in \cite{petri_multi-wavelength_2024} where all the details are given. For completeness, we recall the ingredients in this section, starting with the magnetosphere model, following with the radio and $\gamma$-ray emission sites. We close this section by choosing relevant targets for our samples of MSPs with measured Shapiro delay.

\subsection{Magnetosphere model}

The pulsar magnetosphere, filled with \cor{electron-positron pairs}, is computed in the force-free regime to deduce its magnetic field structure. In this approximation, the fluid characteristics of the plasma are completely ignored, and only its electromagnetic properties are considered. The size of the closed magnetosphere is $\rlight = {c}/{\Omega}$ with $\Omega$ the angular velocity of the pulsar and $c$ the speed of light. Due to computational resource limitations, the ratio between the neutron star radius~$R$ and the light cylinder radius is set to ${R/\rlight=0.1}$ which corresponds to a \cor{pulsar of period 2.5~ms assuming a radius of 12~km}.

\subsection{Radio emission\label{subsec:radio-emission}}

The radio photons originate from the polar caps, defined as the pulsar surface to which are rooted the open magnetic field lines. There are two symmetric polar caps, situated at the magnetic poles of the pulsar. Radio photons are produced through curvature radiation. Because of the high magnetic field, the gyration radius of particles inside the magnetosphere is very small, and it is as if they followed exactly the magnetic field lines. Due to the curved trajectory they adopt, they will emit $\gamma$-ray photons, which cannot escape the magnetosphere. Instead, they produce $e^-e^+$ pairs, that emit again $\gamma$-ray photons through annihilation with another pair, and so on: particles are produced in cascade. An $e^-e^+$ plasma is progressively created, which emits in the radio waveband in the form of a beam.

\subsection{Gamma emission\label{subsec:gamma-emission}}

The discontinuity of the magnetic field outside the light cylinder is at the origin of a current sheet. Due to the rotation of the pulsar, this current sheet will propagate like a "ballerina skirt". The $\gamma$-ray emission emanates from this current sheet and produces a pulsed emission. The number of pulses detected per period depends on the line-of-sight inclination. If the observer looks directly through the current sheet, two pulses per period are observed; if it looks at the edge of the layer, one pulse is seen and if the line-of-sight is outside the current sheet, no pulsation is detected. The emission starts directly at the light cylinder radius, and continues up to several times this radius, with a decaying emissivity. It is assumed that the particles propagate radially at a speed close to the speed of light, thus they will present a focalised relativistic emission. The cone of emission of the particles will depend on the Lorentz factor $\Gamma$ associated to their velocity. We fix $\Gamma=10$ for concreteness.

\subsection{Pulsar geometry\label{subsec:angle-const}}

We define three angles linked to the pulsar geometry in the binary. First we introduce the angle~$\rchi$, which is the angle between the rotation axis and the magnetic axis, also called the obliquity. Next the viewing angle~$\zeta$ represents the angle between the rotation axis and the line-of-sight. Finally, we define the orbital inclination angle~$i$, which is the angle between the line-of-sight and the binary orbital angular momentum. To observe the pulsar shining in $\gamma$-rays, radio, or both, some conditions need to be satisfied. First, for $\gamma$-ray emission, the condition \cor{expressed in radians} reads $|\zeta - \frac{\pi}{2}| \le \rchi$. This ensures that the observer is looking through the current sheet of the pulsar. For radio emission, the ideal configuration corresponds to an observer looking directly at one magnetic pole at least. However, due to the opening of the polar cap, there is a tolerance on the discrepancy between $\zeta$ and $\rchi$. In general, the condition \cor{expressed in radians} to see the radio emission is $|\zeta - \rchi | \le \rho$, with $\rho$ the half-opening angle of the radio beam cone. By symmetry, the condition for the other pole is $|\pi - \zeta - \rchi | \le \rho$.

\subsection{The MSP target sets}


Our approach requires radio-loud $\gamma$-ray MSPs orbiting in a binary system with precise orbital inclination measurements obtained, for instance, from the Shapiro delay. We have selected two MSP samples satisfying this criteria. In a first sample, the angle~$i$ is accurately determined within $1\degr$ or even less whereas for a second sample, it is only loosely constrained with an interval spanning $10\degr$ or more. Cross-matching the 3PC with reported Shapiro delays, we found a small sample of \cor{14} pulsars for the first set listed in Table~\ref{tab:Sample_Shapiro1} and \cor{15} pulsars with credible intervals in table~\ref{tab:Sample_Shapiro2} for the second set. 
The essential features, spin period~$P$, orbital inclination angle~$i$ from Shapiro delay and companion type are \cor{also} given. 
\cor{The $\gamma$-ray light curves for both samples are taken from the Third Fermi/LAT Catalog of Gamma-ray Pulsars Catalog (3PC), in which the radio pulse profiles are also provided.}

\begin{table}[h]
	\caption{Radio-loud $\gamma$-ray MSPs with measured spin period~$P$, orbital inclination angle $i$ from Shapiro delay and companion type (WD for white dwarf, He for Helium and CO for Carbon-Oxygen).\label{tab:Sample_Shapiro1}}
	\centering
	\begin{tabular}{llll}
		\hline
		Name & $P$ (ms) & $i$ (deg) & Companion \\ 
		\hline
		J0218$+$4232  & 2.32 & 85.1 & He WD \\ 
		J0437$-$4715  & 5.76 & 42.42 & He WD \\ 
		J0737$-$3039A & 22.7 & 89.35 & NS \\ 
		J0740$+$6620  & 2.89 & 87.56 & He WD \\ 
		J0955$-$6150  & 1.99 & 83.2 & \\ 
		J1012$-$4235  & 3.1  & 87.97 & He WD\\ 
		J1125$-$6014  & 2.63 & 77.6 & He WD\\ 
		J1600$-$3053  & 3.6 & 68.6 & He WD\\ 
		J1614$-$2230  & 3.15 & 89.18 & CO WD \\ 
		J1713$+$0747  & 4.57 & 72.0 & He WD \\ 
		J1741$+$1351a & 3.75 & 73 & He WD\\ 
		J1857$+$0943  & 5.36 & 88.0 & He WD \\ 
		J1909$-$3744  & 2.95 & 86.69 & He WD\\ 
		J2043$+$1711  & 2.38 & 83.2 & He WD \\ 
		\hline
	\end{tabular}
\end{table}

\begin{table}[h]
	\caption{Radio-loud $\gamma$-ray MSPs with measured spin period~$P$, orbital inclination angle estimates $i$ and companion type. The abbreviations WD and UL stand for white dwarf and ultra-light. \label{tab:Sample_Shapiro2}}
	\centering
	\begin{tabular}{llll}
		\hline
		Name & $P$ (ms) & $i$ (deg) & Companion \\ 
		\hline
		J0023$+$0923 & 3.05 & $54 \pm 14$ & UL \\ 
			& & $77 \pm 13$ & \\ 
			& & $42 \pm 4$ & \\ 
		J0101$-$6422 & 2.57 & 70-75 & He WD \\ 
		J0610$-$2100 & 3.86 & $73.2^{+15.6}_{-19.2}$ & UL \\ 
		J0636$+$5128 & 2.87 & $24.3 \pm 3.5$ & UL \\ 
		& & $23.3 \pm 0.3$ & \\ 
		& & $24.0 \pm 1.0$ & \\ 
		J1124$-$3653 & 2.41 & $44.9 \pm 0.4$ & \\ 
		J1514$-$4946 & 3.59 & 68-82 & He WD \\ 
		J1544$+$4937 & 2.16 & $47^{+7}_{-4}$ & UL \\ 
		J1555$-$2908 & 1.79 & $i > 75$ & UL \\ 
		& & $i > 83$ & \\ 
		J1628$-$3205 & 3.21 & $i>55$ & MS \\ 
		& & $i<82.2$ & \\ 
        J1640$+$2224  & 3.16 & $84^{+4}_{-6}$ & He WD (?) \\ 
         & & $60\pm6$ &  \\ 
		J1732$-$5049 & 5.31 & 59-72 & He WD \\ 
        J1811$-$2405  & 2.7 & $76.2^{+2.8}_{-3.2}$  & He WD \\ 
		J1959$+$2048 & 1.61 & $65 \pm 2$ & UL \\ 
		& & $62.5 \pm 1.3$ &  \\ 
		& &  $85.1 \pm 0.4$ & \\ 
		J2051$-$0827 & 4.51 & $55.9^{+4.8}_{-4.1}$ & UL \\ 
		& & $59.5 \pm 0.4$ & \\ 
		J2256$-$1024 & 2.29 &  $68 \pm 11$ & UL \\ 
		\hline
	\end{tabular}
\end{table}

\section{Fitting of the light curves\label{sec:Method}}


Because \cor{the $\gamma$-ray} light curves are periodic functions of the rotational \cor{phase~$\varphi\in[0,1]$}, in a first stage, we use Fourier transforms to interpolate the predicted $\gamma$-ray profiles~\cor{$\mathcal{M}(\varphi)$} at any phase~$\varphi$ to high accuracy. \cor{In a second stage, we look for the best shift~$\phi$, the best scaling~$a$ and the best background level~$b$ to apply to the modelled signal~$\mathcal{V}(\varphi)$ so that it best approaches the observed signal~$\mathcal{U}(\varphi)$, by writing
\begin{equation}
    \mathcal{U}(\varphi) = a \, \mathcal{V}(\varphi-\phi) + b + \mathcal{N}
\end{equation}
with $\mathcal{N}$ some noise in the data, see Appendix~\ref{appendix:crosscorrelation}.} \cortwo{The optimal shift $\phi$ fixes primarily the degeneracy between two-peak $\gamma$-ray profiles, depending on whether or not the separation between the first and the second peak is larger than half a period. This parameter is also used to correct from possible non-radial emission, not taken into account by our $\gamma$-ray emission model, but only for small phase shifts (smaller than $0.25$ in absolute value). This effect is discussed in more details in section~\ref{sec:discussion-non-radial-prop}.} Finally, to fit our model, we use Bayesian inference to extract the maximum likelihood value and get an estimate of the error in the fitted angles $(\rchi, \zeta)$. \cor{Appendix~\ref{appendix:fittingprocess} gives full details about this method.}


\cor{Before finding these best parameters $(a,b,\phi)$ and performing the fit, we properly phase align the modelled $\gamma$-ray pulses with the radio pulse profile, such that phase zero $\varphi=0$ corresponds to the peak of the radio pulse. Note that this phase alignment is already done for the $\gamma$-ray light curves taken from 3PC. We emphasize that the fitting is not performed on the radio profile. However, once the characteristic angles $(\zeta,\rchi)$ have been obtained through the $\gamma$-ray light curve fitting, we can predict the corresponding radio profile, Sec.~\ref{subsec:radio-emission}. This predicted profile is then used as a visual cross-check of the accuracy of our fit, on top of more formal verifications that are described in section \ref{sec:KS-test}. This visual verification is based on whether or not we recover a possible interpulse and its separation to the main pulse, without accurately looking at the radio pulse shape.}

\subsection{Bayesian inference}

Finding the model that best fits the observed \cor{$\gamma$-ray} light curve is a typical Bayesian inference problem. \cor{The parameters of the fitting being $\theta=(\rchi,\zeta)$, we look for the posterior} of the parameter \cor{$\theta$} and maximise it, in order to obtain the best fit parameters. We compute \cor{the posterior} distribution thanks to Bayes theorem \cor{(see Appendix~\ref{appendix:fittingprocess} for more details)}. The prior for $\theta$ is taken to be uniform in
\cor{\begin{equation}
    \Theta =\{(\chi,\zeta) \in \left[0,\frac{\pi}{2}\right]^2 :  \left|\zeta - \frac{\pi}{2}\right| \le \chi \text{ and } |\zeta - \chi| \le \rho   \}
\end{equation}}
\cor{which guarantees that the pulsar is seen in radio and in $\gamma$-rays, Sec.~\ref{subsec:angle-const}. A tolerance of $10\degr$ is applied to the intervals constrained by those two observational conditions, since the transition is not that sharp but certainly smoother.} 
\cor{As it is described precisely in Appendix~\ref{appendix:fittingprocess}}, maximising the posterior is equivalent to maximise the likelihood, or in the same way its logarithm. By taking a Gaussian distribution for the likelihood, we recover the least square method. In practice, the computation, \cor{including uncertainties,} is made using the Python library Bilby, which is a Markov chain Monte Carlo \cor{(MCMC)} sampling algorithm, originally implemented for the analysis of gravitational waves from merging compact objects \citep{ashton_bilby_2019}. The error bars are taken at a $3\sigma$ confidence interval.
After performing the fitting of all the pulsars \cor{$\gamma$-ray} light curves from our sample, we compare the best \cor{line-of-sight angle} $\zeta$ fit to the inclination angle $i$ provided by Shapiro delay measurements. \cor{As a check of the best fit,} we perform a second fitting of all the \cor{$\gamma$-ray} light curves, but now imposing an \cor{almost} perfect alignment, meaning $\zeta=i$. \cor{For this purpose,} the prior for $\zeta$ is now taken to be a Gaussian of mean~$i$ and variance $\sigma^2=0.01$, and the prior for $\rchi$ \cor{is} the same as previously.

\subsection{Symmetry property in the $\gamma$-ray emission}

\cor{The model used for the $\gamma$-ray light curves predicts a symmetry in the $\rchi$-$\zeta$ plane with respect to the line $\rchi=\zeta$. This means that for a solution ($\rchi_1,\zeta_1$), there should also exist a symmetrical solution ($\rchi_2=\zeta_1, \zeta_2=\rchi_1$). This comes from the following formula, which predicts the separation between the pulses \citep{petri_unified_2011}}
\cor{\begin{equation}
    \cos(\pi\Delta)= |\cot(\zeta)\cot(\rchi)|
\end{equation}}
\cor{with $\Delta$ the peak separation. It can be seen that, by inverting the values of $\rchi$ and $\zeta$, the separation remains the same. The symmetry is not exact for a dipolar model, thus mathematically only one solution will maximize our likelihood (or in the same way minimize the $\rchi^2$ from the least square method) during the Bayesian inference process. However, in our study, the second symmetrical solution might be physically more relevant, if it favours alignment. The selection method we applied is the following. We run the inference on the whole parameter space. If the absolute best fit found is in agreement with the hypothesis of alignment, we keep this solution. If it is not, we perform a second fit using the symmetrical parameter space in order to find a second local minimum, and we test the hypothesis of alignment on this solution. If the value of its $\rchi^2$, written $\rchi^2_{\rm min'}$, is considered as equivalent to the one of the first solution, $\rchi^2_{\rm min}$, then we keep this solution for our analysis. By equivalent, it is meant that the value of $\rchi^2_{\rm min'}$ is within an isocontour at $\rchi^2_{\rm min} +2.71$, which corresponds to a confidence interval at $90\%$ \citep{press_numerical_2007}.}

\subsection{Test of the fitting quality}\label{sec:KS-test}

In order to assess the quality of the fitting, a Kolmogorov-Smirnov (KS) test is performed on the distribution of the residuals $\{R_n\}$ ($n$ being the number of data points) defined by \citep{andrae_dos_2010} $R_n = {(y_{{\rm data},n}-y_{{\rm model},n})}/{\sigma_n}$, with $y_{{\rm data},n}$ the data \cor{set, $y_{{\rm model},n}$ the predicted set} after the fit, and $\sigma_n$ the uncertainty on the data points. If the fit is accurate, the distribution of the residuals should follow a Gaussian of mean $\mu = 0$ and variance $\sigma^2 = 1$: this is the null hypothesis. \cor{The Kolmogorov–Smirnov test compares the distribution of the residuals to a Gaussian distribution to quantify how much they differ. The test statistic $D = {\rm max} |F_1(x) - F_2(x)|$, measures the maximum distance between two cumulative distributions $F_1$ and $F_2$. In our case, $F_1$ would be the Gaussian distribution and $F_2$ the distribution of the residuals. From this statistic, a p-value is computed. If the p-value is larger than \cor
{a threshold $\alpha$ usually fixed to $\alpha=5$\%}, meaning that the null hypothesis cannot be rejected at a $5$\% confidence level, the two distributions are close enough to consider the fit accurate. This test is still not entirely reliable, and we will see that for some pulsars of the sample we will need to qualify its result.}

\section{Fitting results\label{sec:Results}}

\cor{We separate our results in two parts corresponding to our two samples of MSPs. The first sample belongs to binary systems for which the orbital inclination~$i$ is known accurately. The second sample belongs to binary systems for which the orbital inclination~$i$ is less well constrained. For both sample, the solution considered by default is the one of the first fit, and if the symmetrical solution has been kept it will be stated in the description of the corresponding pulsar. For those pulsars, the original solution with the absolute minimum is shown in \ref{appendix:first-fit}.}

\subsection{First sample}

\begin{figure*}[h]
	\centering
	\includegraphics[width=\linewidth]{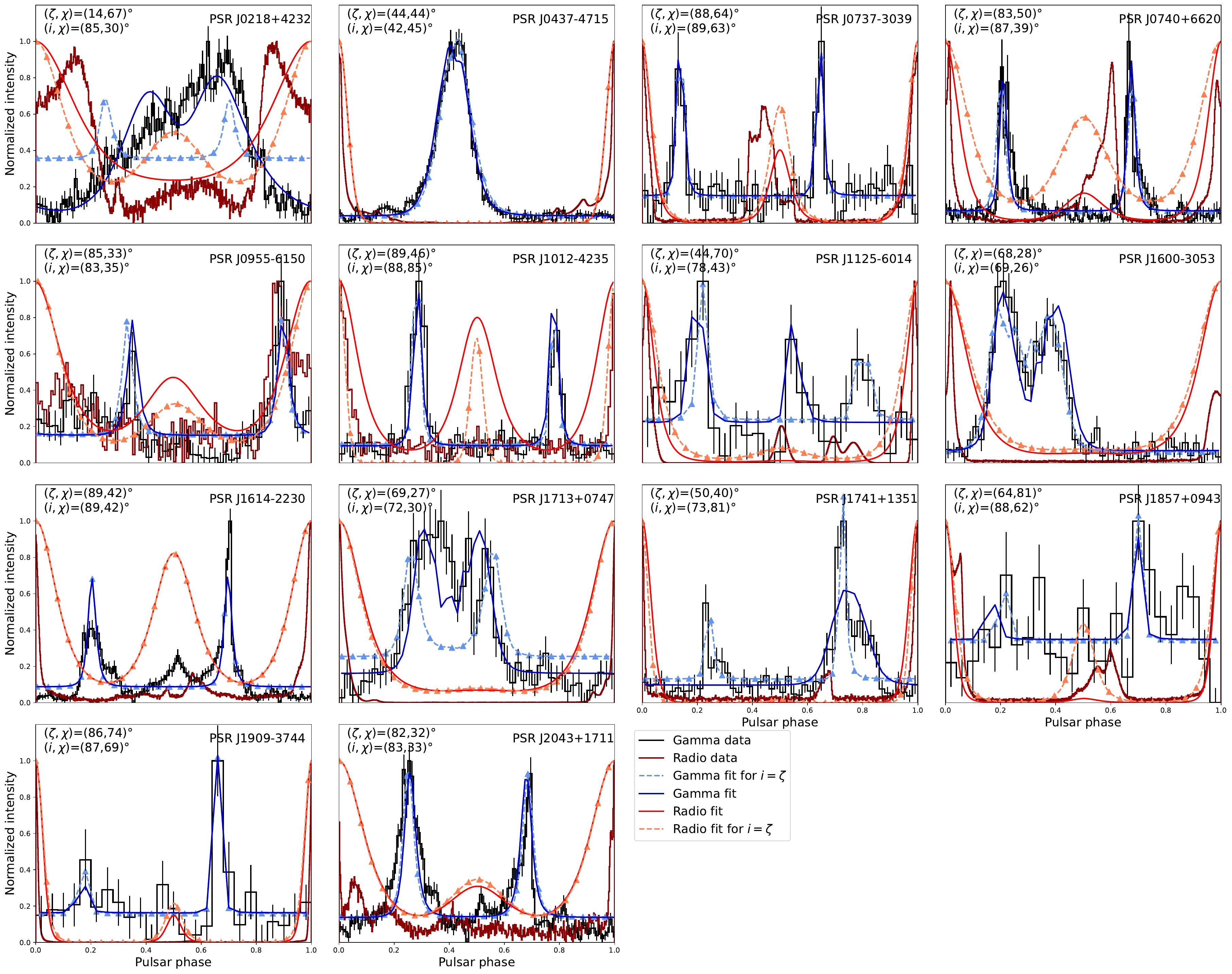}
	\caption{Fitting of the light curves from the first sample of MSPs, with the \cor{black} curve and the dark red curves corresponding respectively to the $\gamma$-ray and radio data, the light blue and the orange curves corresponding respectively to the $\gamma$-ray light curves and radio profiles, finally the light blue dotted and orange dotted curves obtained from the same fitting but imposing a perfect alignment, meaning $\zeta=i$.}
	\label{fig:MSP1-panel-result}
\end{figure*}

We start with the first sample of MSP. Figure~\ref{fig:MSP1-panel-result} summarises the $\gamma$-ray light curve fitting for the whole sample. The \cor{black} and dark red curves correspond respectively to the $\gamma$-ray and radio data; the blue and the \cor{red} curves correspond respectively to the predicted $\gamma$-ray light curves and radio pulse profiles \cor{plotted according to characteristic angles obtained after the fitting of the $\gamma$-ray light curve}. Finally, the light blue and orange dotted curves \cor{with triangles} are obtained from the same fitting but imposing perfect alignment with \cor{the constrain} $\zeta=i$. A description of those curves for each pulsar individually is given in the following paragraphs. 


\subsubsection{\cor{J0218+4232}}


This pulsar of period $P=\SI{2.32}{\milli\second}$ shows one wide $\gamma$-ray peak during almost the whole period, and one radio peak with a complex shape, \cor{which could be due to a hollow cone geometry,} or possibly characteristic of multipolar magnetic field components. The radio profile also presents a wide interpulse with a small amplitude. The light curve fitting provides $\rchi=66.6 [+0.4,-0.6]\degr$, $\zeta= 14.2 [+1.7,-1.1]\degr$ and $\phi=0.095$. Rather than one wide peak, the modelled light curve shows two unresolved $\gamma$-ray peaks. Indeed, this type of light curve with one very large peak is not well predicted by the model. However, the fitting is still accurate, with a p-value of $0.19$. The predicted radio profile is also rather good, considering the complex shape of the radio pulse, however we miss the small interpulse. The fitted $\zeta$ angle is far from what has been obtained with Shapiro delay, with $i=85.2\degr$ \citep{tan_high-cadence_2024}. When a second fit imposing perfect alignment is done, the magnetic angle obtained is $\rchi_i=30.0 [+0.9,-0.1]\degr$ \cortwo{with a phase shift of $\phi_i=0.035$}. The curve deviates even further from the observed light curve, with a p-value of $0.0005$. Indeed, we can see that the corresponding $\gamma$-ray light curve has two well resolved pulses, which does not match at all the observed light curve. However, the radio pulse profile is rather accurate, showing both the pulse and the interpulse. With those results, we can conclude that this pulsar does not satisfy the hypothesis of alignment.


\subsubsection{\cor{J0437-4715}}
This pulsar of period $P=\SI{5.76}{\milli\second}$ shows one $\gamma$-ray peak and one radio peak. The best fitting parameters for are $\rchi=44.4 [+3.5,-12.8]\degr$, $\zeta=44.0 [+12.3,-3.5]\degr$ and $\phi=-0.04$, with a p-value of $0.82$. This is consistent with the orbital inclination angle deduced from Shapiro delay with $i'=42.42\degr$, assuming the complementary angle, $i'=\pi-i$ which gives the same $\gamma$-ray light curves due to symmetry considerations \citep{reardon_timing_2016}. When a second fit imposing a perfect alignment is performed, we indeed recover a very similar fitting, with a magnetic angle $\rchi_i=44.5[+3.4,-0.5]\degr$ \cortwo{and a phase shift $\phi_i=-0.035$}, as well as a slightly higher p-value of $0.91$.

\subsubsection{\cor{J0737-3039A}}
This pulsar has the largest period of the sample, with ${P=\SI{22.7}{\milli\second}}$. It belongs to a very peculiar system, the only double pulsar system known so far. It shows two $\gamma$-ray peaks and a radio peak with interpulse, hinting for an almost perpendicular rotator. The fitting of the $\gamma$-ray light curve provides $\rchi=63.7 [+26.1,-8.9]\degr$, $\zeta=88.2 [+1.8,-18.1]\degr$ and $\phi=0.45$. \cortwo{The separation of the two peaks being $\Delta=0.506\pm0.012$, the large phase shift can be attributed to an inversion of the peaks.} The fit is accurate, with a p-value of $0.93$. This is coherent with the Shapiro delay, with $i=89.35\degr$ \citep{kramer_strong-field_2021}. The resulting modelling of the radio pulse profile gives a radio interpulse with an amplitude and a shape that is not exactly the one obtained through observations, but it is well centred on the observed peak, which is what we are looking for with this model. When a second fit imposing a perfect alignment is done, we can see that we recover a very similar fitting, with a magnetic angle $\rchi_i=63.3 [+24.7,-6.8]\degr$ \cortwo{and the same large phase shift $\phi_i=0.45$}, and a p-value of $0.93$.

\subsubsection{\cor{J0740+6620}}


This pulsar of period $P=\SI{2.89}{\milli\second}$ shows double $\gamma$-ray peaks and a radio peak with an interpulse. The results of the fit are $\rchi= 50.1 [+8.8,-7.1]\degr$, $\zeta = 83.4 [+2.2,-3.2]\degr$, and $\phi= -0.005$. The fitting of the $\gamma$-ray light curve is accurate, with a p-value of $0.41$. The modelled radio profile recovers the interpulse, however, with an amplitude that is too small. The fitted $\zeta$ angle is only a few degrees away from the inclination given by Shapiro delay, with $i=87.56\degr$ \citep{fonseca_refined_2021}. For the second modelling, imposing a perfect alignment, we recover a radio interpulse with a larger amplitude. The magnetic angle found is $\rchi_i=38.8 [+2.9,-0.9]\degr$, \cortwo{with a phase shift $\phi_i=-0.005$}. The $\gamma$-ray light curve obtained is similar to the previous one, leading to a p-value of $0.16$. 



\subsubsection{\cor{J0955-6150}}
This pulsar of period $P=\SI{1.99}{\milli\second}$ is the fastest pulsar of the sample. It shows one strong and one weak $\gamma$-ray peak, as well as a radio peak with a complex shape. For this pulsar, the symmetrical solution has been taken. The fitting gives $\rchi=32.6 [+10.5,-9.6]\degr$, $\zeta= 85.2 [+1.7,-9.2]\degr$ and $\phi=-0.31$. Due to the noise and the large delay between the two $\gamma$-ray peaks, the fitting was hard to perform, only the right part of the first $\gamma$-ray peak is recovered and the second $\gamma$-ray peak is too thin and has an amplitude too small compared to what is observed. \cortwo{Since the radio pulse is quite wide, it is hard to define the phase $0$ of this profile. This uncertainty could explain the large phase shift obtained for this pulsar, better than a non-radial propagation effect in the $\gamma$-ray emission.} The predicted radio profile shows a radio interpulse that is not observed, and overestimates the signal between the peaks. The fit is still accurate, with a p-value of $0.43$. The best fit $\zeta$ angle is consistent with Shapiro delay measurements, with $i=83.2\degr$ \citep{serylak_eccentric_2022}, the two fits being very similar, only the first $\gamma$-ray peak is a bit shifted to the left. The magnetic angle found for the fit imposing the Shapiro constraint is $\rchi_i=34.9[+15.9,-6.0]\degr$, \cortwo{the phase shift is the same as previously with $\phi_i=-0.31$}. The resulting p-value obtained is $0.51$, thus it is still accurate. The predicted radio pulse profile is slightly more accurate than the previous one, with a smaller interpulse.

\subsubsection{\cor{J1012-4235}}
This pulsar of period $P=\SI{3.1}{\milli\second}$ shows a double peaked $\gamma$-ray profile and one radio peak. For this pulsar, the symmetrical solution has also been taken. The fitting of the $\gamma$-ray light curve gives $\rchi= 46.3 [+13.7,-4.2]\degr$, $\zeta= 89.4 [+0.6,-2.0]\degr$, and $\phi= 0.09$. The fit is accurate, with a p-value of $0.58$. The modelled radio pulse profile shows an interpulse that is not observed. The reason for this is given just below. The $\zeta$ angle obtained is compatible with the inclination given by Shapiro delay~\citep{gautam_detection_2024}, $i=87.97\degr$. When a second fit imposing a perfect alignment is done, the $\gamma$-ray light curve is thus very similar to the previous one, \cortwo{but this time} with a \cortwo{larger} magnetic angle $\rchi_i=85.0[+4.9,-10.7]\degr$. \cortwo{This however does not impact the shape of the modelled light curve, due to symmetries in the $\gamma$-ray emission model. The phase shift is now also much larger, with $\phi_i=-0.41$. The separation between the peaks could be larger than $0.5$ considering the error bars, with $\Delta=0.496\pm0.008$. Thus having a best fit which requires an inversion of the peaks is mathematically still possible, however since there is no observed radio interpulse we do not have another way to define the beginning of the period and justify the inversion of the peaks (see Sec \ref{sec:discussion-non-radial-prop} for more details).} The p-value is now of $0.39$. The radio pulse profile is slightly better than the previous one, still predicting an interpulse but slightly smaller than before.  

The model assumes that the polar caps, from which are emitted the radio photons, are antipodal, but recent observations from the NICER mission \citep{gendreau_neutron_2012} tend to show that this is not necessarily true. If this is the case, radio photons for an aligned rotator could be observed by looking at its equator, or could miss an interpulse that should be observed for an orthogonal rotator. This effect is particularly important for MSPs, for which radio and $\gamma$-ray emission occurs close to the stellar surface. Since this feature is not taken into account in our model, this could explain the possible missing or false prediction of a non-existing interpulse in the radio profile resulting from an accurate $\gamma$-ray fit.

\subsubsection{\cor{J1125-6014}}
This pulsar of period $P=\SI{2.63}{\milli\second}$ shows  weak double $\gamma$-ray peaks, and a radio peak with a weak interpulse. The number of data points for this pulsar being smaller than $30$, using a KS test is less relevant. The fitting obtained seems rather good considering the noise in the data, even though the second $\gamma$-ray peak shows a complex structure possibly associated to a third peak, the fit focusses on the left part of the pulse. The predicted radio profile recovers well the main pulse, but misses the radio interpulses, probably for the same reason as for PSR~J1012-4235. The best fitting parameters are $\rchi= 69.8 [+14.9,-47.2]\degr$, $\zeta= 44.1 [+38.5,-15.8]\degr$ and $\phi= -0.06$. The best fit $\zeta$ angle is far from the value given by Shapiro delay, $i=77.6\degr$ \citep{shamohammadi_searches_2023}, but given the error bars, alignment is not to exclude. When a second fit imposing a perfect alignment is done, we recover a similar fitting : the first $\gamma$-ray peak is slightly thinner than the previous one, and the second one \cortwo{now rather focusses on the right part of the possible double pulse.}. The magnetic angle obtained for this second fit is $\rchi_i=42.7[+44.1,-13.4]\degr$, \cortwo{and due to the shift of the second fitted peak the phase shift is now much larger, with $\phi_i=-0.42$. Being also a faint pulsar with a low photon statistics, it is hard to draw a firm conclusion}.

\subsubsection{\cor{J1600-3053}}
This pulsar of period $P=\SI{3.6}{\milli\second}$ shows really close double $\gamma$-ray peaks and one radio peak. The results of the fit are $\rchi= 28.2 [+12.1,-3.8]\degr$, $\zeta= 67.7 [+2.9,-11.0]\degr$, and $\phi= -0.13$. The fit is accurate, with a p-value of $0.75$. The modelling of the radio profile is accurate, even though it slightly overestimates the signal between the peaks. The fitted $\zeta$ angle is very close to the measurement deduced from Shapiro delay, with $i=68.6\degr$ \citep{desvignes_high-precision_2016}. Performing a second fit by imposing a perfect alignment, we recover very similar fitting parameters, both in $\gamma$-rays and in radio, with a magnetic angle $\rchi_i=25.7[+1.3,-0.7]\degr$, \cortwo{a phase shift $\phi_i=-0.15$} and a p-value of $0.82$.

\subsubsection{\cor{J1614-2230}}
This pulsar of period $P=\SI{3.15}{\milli\second}$ shows two main $\gamma$-ray peaks and one weak peak, as well as one radio peak with a weak interpulse. It is unclear if the weak $\gamma$-ray pulse should be associated to the strong peak or not. The results of the fitting are $\rchi= 41.8 [+0.2,-0.8]\degr$, $\zeta= 89.1 [+0.5,-1.1]\degr$ and $\phi= 0.005$. The p-value is $0.02$, smaller than the threshold fixed at $0.05$ for a $5\%$ confidence interval. This could come from the first $\gamma$-ray peak, that has an amplitude slightly too large compared to what is observed, and the second $\gamma$-ray peak whose amplitude is a bit too small. Nevertheless, the p-value is still close from the threshold value and would be valid considering a criteria at $1\%$ confidence level. The first radio peak is well recovered, but the modelled radio interpulse has a larger amplitude than what is observed. Shapiro delay gives an inclination of $i=89.18\degr$ \citep{shamohammadi_searches_2023}, which is consistent with what has been found for the fitted $\zeta$ angle. Thus when a second fit imposing a perfect alignment is done, we recover a very similar fitting, with almost the same magnetic angle, $\rchi_i=41.7[+0.3,-0.3]\degr$, \cortwo{a phase shift $\phi_i=0.005$} and the same p-value of $0.02$. This low p-value should be attributed to the fact that the third weak peak $\gamma$-ray peak is not reproduced by the model. It shows some similarities with PSR~J1125-6014.

\subsubsection{\cor{J1713+0747}}
This pulsar of period $P=\SI{4.57}{\milli\second}$ shows one large $\gamma$-ray peak spread over a significant fraction of the period, and one radio peak. As for PSR~J0218+4232, this type of $\gamma$-ray light curve is not well reproduced by our model. The symmetrical solution has been kept, as for PSR~J0955$-$6150 and PSR~J1012$-$4235. The best fit parameters are $\rchi= 26.7 [+2.3,-6.4]\degr$, $\zeta= 69.1 [+0.7,-18.0]\degr$, and $\phi=-0.05$. The fitting is still accurate with a p-value of $0.64$. The fitted $\zeta$ angle is only a few degrees away from what is given by Shapiro delay, with $i=72 \degr$ \citep{fonseca_nanograv_2016}. When a second fit imposing a perfect alignment is done, we obtain two unresolved $\gamma$-ray peaks rather than one wide peak. The magnetic angle found is $\rchi_i=29.6[+0.4,-0.5]\degr$ \cortwo{with a phase shift $\phi_i=-0.03$}, and the p-value is now of $0.24$. The radio pulse profile is however very similar.

\subsubsection{\cor{J1741+1351a}}
This pulsar of period $P=\SI{3.75}{\milli\second}$ shows one weak and one strong $\gamma$-ray peak, as well as one radio peak with a weak interpulse. The fitting parameters are $\rchi= 39.8 [+13.0,-8.6]\degr$, $\zeta= 50.1 [+8.6,-11.4]\degr$, and $\phi= 0.31$. The fit is accurate, with a p-value of $0.30$, even if we completely miss the first $\gamma$-ray peak. Indeed, this peak is not statistically significant since it is mostly composed of one small amplitude data point. \cortwo{The large phase shift can also be attributed to the missing $\gamma$-ray peak. The modelled $\gamma$-ray light curve only reproduces one peak, but since the actual light curve shows two peaks, the predicted single peak light curve has to be shifted to later phases to fit the second observed peak.} Concerning the radio profile, we miss the small interpulse. The inclination given by Shapiro delay slightly differs from the best fit $\zeta$ angle, even considering the error bars, with $i= 73\degr$ \citep{arzoumanian_nanograv_2018}. When a second fit imposing a perfect alignment is done, we indeed recover two peaks, but the second one remains slightly too thin compared to the observations. The magnetic angle found is $\rchi_i=81.2[+0.8,-1.2]\degr$ \cortwo{with a phase shift $\phi_i=0.05$}, and the p-value becomes $0.44$. The predicted radio profile shows a very small interpulse, not centred on the one observed. Even if both fits (general and imposing alignment) pass the KS test, physically it is more accurate to find two $\gamma$-ray pulses, which favours the solution with the Shapiro constraint, so we do not have enough arguments to exclude the hypothesis of alignment for this pulsar.

\subsubsection{\cor{J1857+0943}}
This pulsar of period $P=\SI{5.36}{\milli\second}$ shows a weak signal, made of double $\gamma$-ray peaks and one radio peak with an interpulse. The number of data points for this pulsar is smaller than $30$, making a KS test less relevant and the peak structure not particularly visible. Even if considering the noise in the data, the fit is overall reasonable, the first peak is poorly reproduced, its amplitude is very small, and it does not seem well centred. The second peak amplitude is a better, but it is too thin and not well centred. The predicted radio profile reproduces well the main pulse, but misses the interpulse. It gives $\rchi= 80.9 [+9.1,-52.6]\degr$, $\zeta= 63.8 [+26.1,-35.8]\degr$ and $\phi= 0.50$. \cortwo{Given the large uncertainties, it is hard to reach any conclusion concerning the large phase shift.} The inclination given by Shapiro delay is $i=88.0\degr$ \citep{arzoumanian_nanograv_2018}, which, considering the error bars, is in agreement with the value of the best fit $\zeta$ angle. When a second fit imposing a perfect alignment is done, the $\gamma$-ray light curve obtained is slightly different from the one found previously. The first $\gamma$-ray peak obtained seems better centred on the observed one, but is thinner. This time we recover a small interpulse for the radio profile. The magnetic angle found for this second fit is $\rchi_i=62.5[+27.4,-16.4]\degr$ \cortwo{and the phase shift is $\phi_i=0.02$}. The statistics of this pulsar is certainly too poor to make any firm conclusion about the fitting results.

\subsubsection{\cor{J1909-3744}}
This pulsar of period $P=\SI{2.95}{\milli\second}$ shows weak double $\gamma$-ray peaks and one radio peak. The number of data points for this pulsar is smaller than $30$, making a KS test less relevant. However, the fitting of the $\gamma$-ray light curve seems accurate, even if it is hard to tell if the first $\gamma$-ray peak is well centred or not on the observed pulse. The modelling of the radio pulse profile shows an interpulse that is not observed, the reason is the same as for PSR J1012$-$4235 and PSR J1125$-$6014 for which we had the same problem. The best fitting parameters are $\rchi= 74.4 [+15.6,-30.9]\degr$, $\zeta= 86.1 [+3.9,-46.6]\degr$, and $\phi= -0.02$. The fitted $\zeta$ angle is consistent with the Shapiro delay expectation, with $i=86.69\degr$ \citep{shamohammadi_searches_2023}. When a second fit imposing a perfect alignment is done, the $\gamma$-ray light curve and radio pulse profile obtained are close to the previous ones, only the amplitude of the first $\gamma$-ray peak is slightly larger compared to the previous result. The magnetic angle found for this second fit is $\rchi_i=69.0[+20.9,-23.7]\degr$ \cortwo{with a phase shift $\phi_i=-0.02$}.

\subsubsection{\cor{J2043+1711}}
This pulsar of period $P=\SI{2.38}{\milli\second}$ shows double $\gamma$-ray peaks and one radio peak. The fit gives $\rchi= 31.8 [+0.9,-0.4]\degr$, $\zeta= 82.2 [+0.4,-0.2]\degr$, and $\phi= 0.02$. The fitting seems to be accurate, however the p-value obtained is of $0.01$, out of the $5\%$ confidence interval but surprising since the fit seems to reproduce well the observations in the $\gamma$-rays. However, the modelled radio profile overestimates the signal between the peaks, and shows an interpulse that is not observed. As for PSR J1614$-$2230, the p-value is still close from the threshold value and would be valid considering a criteria at $1\%$ confidence level. The fitted $\zeta$ angle is very close to the inclination found through Shapiro delay, with $i=83.2\degr$ \citep{fonseca_nanograv_2016}. When a second fit imposing a perfect alignment is done, we find for the magnetic angle $\rchi_i=32.7[+0.3,-0.3]\degr$ \cortwo{and for the phase shift $\phi_i=0.03$}. The result becomes more accurate, with a p-value of $0.13$. The $\gamma$-ray light curve obtained is very similar to the previous one, as well as the radio profile.

\subsection{Discussion\label{subsec:discussionfirstsample}}

\begin{figure*}[h]
	\centering
	\includegraphics[width=1\linewidth]{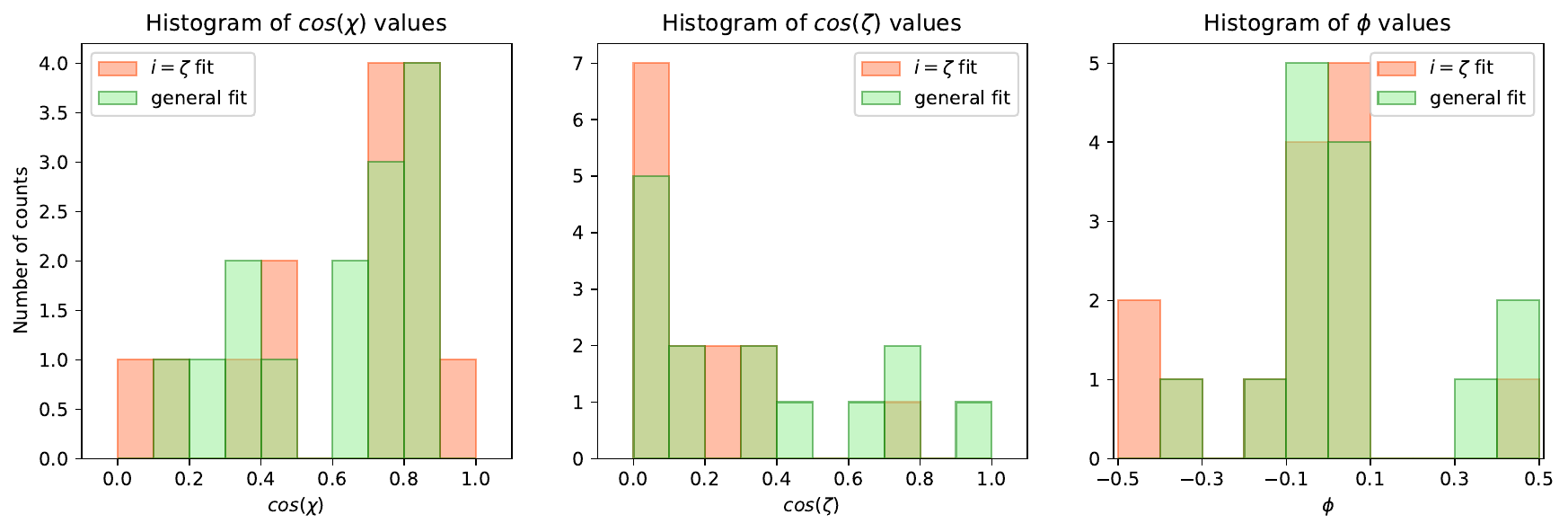}
	\caption{Histogram of best fitting parameters \cor{$\cos\rchi$} (on the left), \cor{$\cos\zeta$} (in the middle), $\phi$ (on the right) for the first sample of MSPs, in green in the general case and in orange for the fitting imposing $i=\zeta$.}
	\label{fig:MSP1-histogram}
\end{figure*}

\cor{Let us summarise our findings from this first sample by showing the histograms of the best fitting parameters $(\rchi,\zeta,\phi)$ in figure~\ref{fig:MSP1-histogram}, in the form of cosine distributions for the angles $\rchi$ and $\zeta$}. Considering the unconstrained fits, the distribution of $\cos \zeta$ is \cor{bimodal}, with one peak close to \cor{$\cos 90\degr = 0$}, whose origin \cor{lies in the selection bias of} our MSP sample. Indeed, one important criteria for their selection was to look for existing Shapiro delay measurements to determine the \cor{orbital} inclination angle~$i$. This Shapiro effect is particularly strong for edge-on binaries, \cor{meaning $i\approx 90\degr$}, which explains \cor{the observed} distribution for $\cos \zeta$. \cor{This bias} is even more pronounced for the distribution of $\cos i$ angles. \cor{However there is also a second peak around $\cos \zeta = \cos 40\degr \approx 0.8$, which is mostly due to the non-aligned pulsars of our sample.} 
The distribution \cor{of $\cos \rchi$} for the same unconstrained fits \cor{peaks at values close to $\cos 40\degr \approx 0.8$, although the statistic is low. The distribution for the constrained fit imposing $i=\zeta$ follows closely the same trend}. However, the sample of MSPs we considered is rather small, so those claims should be taken \cor{with} caution. Considering the distribution of the phase shift~$\phi$, it peaks at a mean value of $0.06$ and has a median of $0.0$, \cortwo{however we can see that a non negligible part of the pulsars from the sample present a large phase shift. A more detailed discussion about this observation is presented in section \ref{sec:discussion-non-radial-prop}, including pulsars from the second sample.}
%

\begin{figure*}[h]
	\centering
	\includegraphics[width=\linewidth]{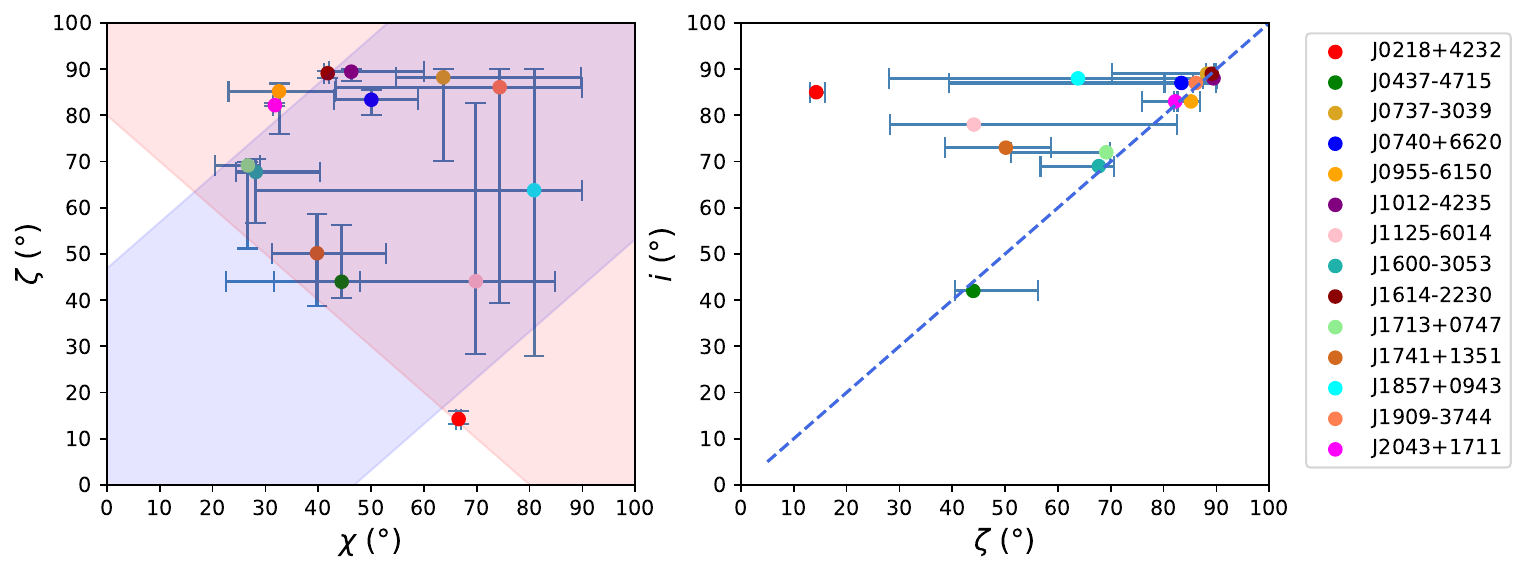}
	\caption{On the left, \cor{the} $\rchi-\zeta$ plane showing the best fitting couple of angles $\rchi$ and $\zeta$ for the first sample of MSPs. On the right, the $\zeta-i$ plane showing the best fitting angle $\zeta$ for the inclination $i$ imposed by Shapiro delay measurements, with the blue dotted line corresponding to the \cor{perfect alignment condition $i=\zeta$}. \cor{Individual pulsars are depicted by different colors.}}
	\label{fig:MSP1-chi-i-zeta-plane}
\end{figure*}

Figure~\ref{fig:MSP1-chi-i-zeta-plane} gives a summary of the best fitting parameters, on the left, in the $\rchi-\zeta$ plane showing the couple $\rchi$ and $\zeta$, and on the right, in the $\zeta-i$ plane. Giving the time scale of alignment during the accretion phase of the binary, we expect those two angles to be equal, as it is represented by the blue dotted line. 
The majority of the pulsars are well aligned on the $\zeta=i$ line : \cor{only one of them, PSR~J0218$+$4232, is completely misaligned, and three of them are only a few degrees away from alignment (between $1\degr$ and $3\degr$) : PSR~J0740$+$6620, PSR~J1713$+$0747, PSR~J2043$+$1711. In the end, $71\%$ of them are perfectly aligned, and we reach $93\%$ of alignment by including those only a few degrees away from perfect alignment. A possible explanation for the misalignment of PSR~J0218$+$4232 could be a movement of precession induced by additional torques acting during the accretion phase, not taken into account by the current model or simply a too long alignment timescale. However the shape of its light curve is not well explained by the model used, thus the quality of the fit can still be questioned. }



In order to verify if the \cor{$93$\%} of aligned pulsars could have been obtained by chance, we performed a reduced $\rchi_{\rm red}^2$ test for the ratio between the viewing angle~$\zeta$ and inclination angle~$i$ following the idea of \cite{laycock_pulse-profile_2025}. The quantity to test is $R={\zeta}/{i}$, which in theory is equal to one, according to our understanding of the evolution of a pulsar in a binary. The associated reduced $\rchi_{\rm red}^2$ is defined by 
\begin{equation}\label{key}
	\rchi_{\rm red}^2 = \frac{1}{d} \sum_{k=1}^{n} \frac{(R_k - 1)^2}{\sigma_{\!k}^{2}}
\end{equation}
$n$ being the number of pulsars in our sample (here $n=16$), $d$ the number of degrees of freedom ($d=n-1$), and $\sigma_{\!k}$ the error on the ratio $R$, obtained from the propagation of the errors made on the estimation of $\zeta$. Its value must be compared to a critical value associated to a probability to find by chance a result above this critical value. The latter is found in a $\rchi_{\rm red}^2$ table, and for a confidence level of $99.9$\% (meaning having a probability of $0.001$ to find a higher value), the critical value is $\rchi^2_{\rm crit} \approx 2.45$. However the value for our sample is \cor{$\rchi_{\rm red}^2=49.15 \gg \rchi^2_{\rm crit}$}, far above the critical value, and thus excluding the possibility that this result has been obtained by chance.

\subsection{Second sample}

\begin{figure*}[h]
    \centering
    \includegraphics[width=\linewidth]{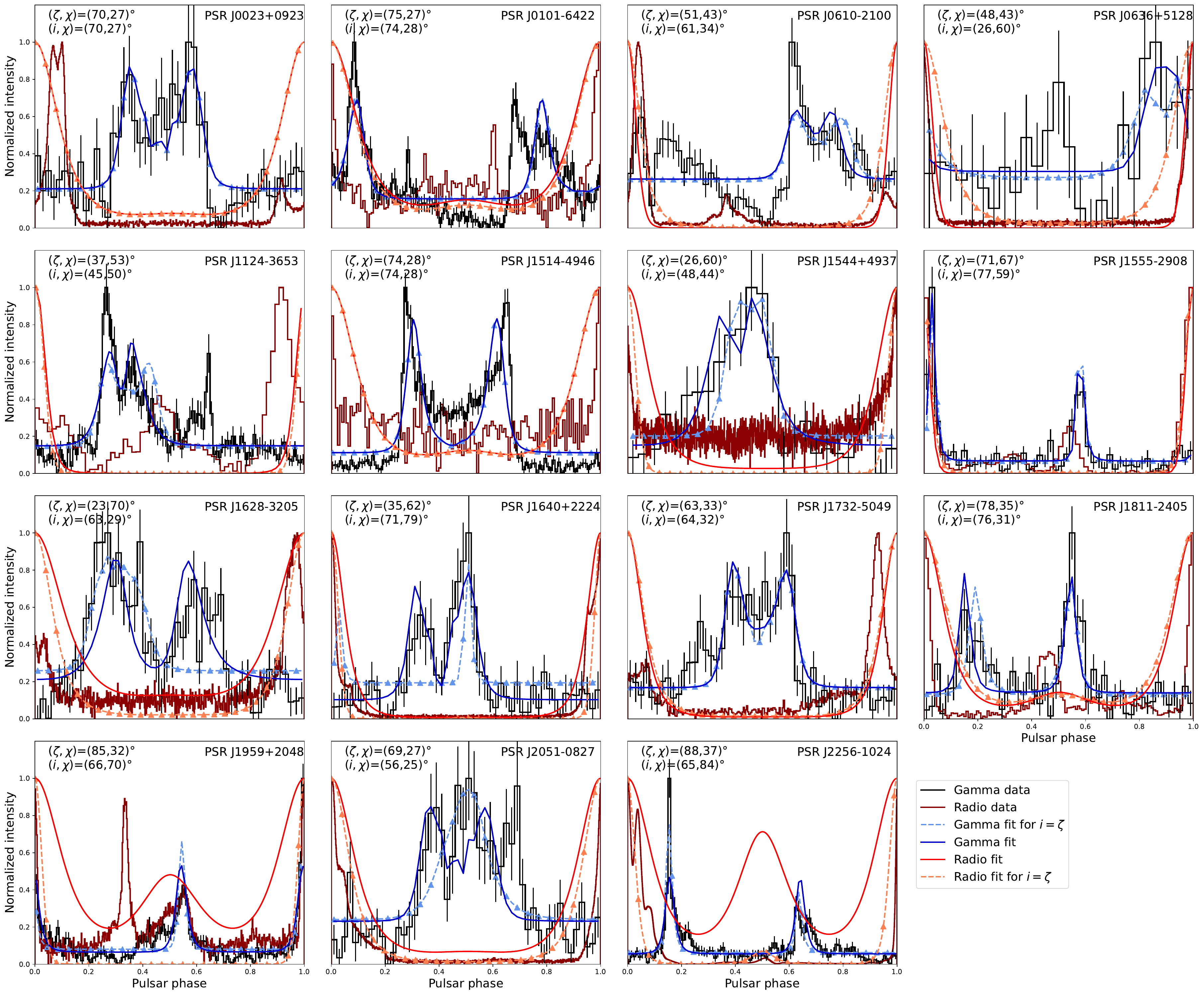}
    \caption{\cor{Same as in Fig.\ref{fig:MSP1-panel-result} but for} the second sample of MSPs.} 
    \label{fig:MSP2-panel-result}
\end{figure*}

The second sample of MSPs is based on \cite{blanchard_census_2025}, who studied the phenomenon of eclipses in spider binary pulsars. For the sample they considered, they provide the different inclination angle measurements available in the literature. Those are summarised in Table~\ref{tab:Sample_Shapiro2}. The idea is to use the same algorithm as previously to fit the $\gamma$-ray light curves of this new sample of MSPs, in order to left the degeneracy between those results. The fitting is done a first time without any constraints on the viewing angle $\zeta$, and then a second time with a restriction of the possible values of this angle on the interval given by the literature in Table~\ref{tab:Sample_Shapiro2}. The different fits obtained are shown on figure \ref{fig:MSP2-panel-result}, using the same legend as in figure~\ref{fig:MSP1-panel-result}. The quality of the fits is discussed in the following paragraphs. 

\subsubsection{\cor{J0023+0923}} 
This pulsar of period $P=\SI{3.05}{\milli\second}$ \citep{breton_discovery_2013, draghis_multiband_2019, mata_sanchez_black_2023} shows two unresolved $\gamma$-ray peaks, and one complex radio peak with three components. The parameters of the fit are $\rchi=27.0 [+11.5,-3.1]\degr$, $\zeta=70.1 [+2.9,-10.1]\degr$, and $\phi=0.01$. The $\gamma$-ray fit is accurate, with a p-value of $0.93$. The predicted radio profile does not reproduce in details the complex structure of the radio peak, but once again this is not what we are looking at. For the second fit made on the interval for the inclination given in \cite{blanchard_census_2025}, the best fit angles are $\rchi_i=27.2 [+31.2,-3.6]\degr$, $i=70.0 [+2.9,-28.8]\degr$, and $\phi_i=0.03$. The value of the inclination is compatible with the hypothesis of alignment. This second $\gamma$-ray fit is still accurate, with a p-value of $0.85$, and it is very similar to the previous one, both in $\gamma$-ray and in radio.

\subsubsection{\cor{J0101-6422}} 
This pulsar of period $P=\SI{2.57}{\milli\second}$ shows one strong peak and one weak interpulse in the radio and double peaks in the $\gamma$-ray domain. For this pulsar, the symmetrical solution has been taken. The $\gamma$-ray light curve and radio pulse profile are rather well fitted, even though the amplitude of the first modelled $\gamma$-ray peak is small compared to observations. The result is $\rchi=26.6 [+1.1,-0.6]\degr$, $\zeta=75.4 [+0.6,-0.4]\degr$ and $\phi=0.495$ for the best fit. \cortwo{The separation of the two peaks is smaller than $0.5$ considering the maximum of the peaks, however the second predicted $\gamma$-ray peak is not centred on a maximum but is shifted to the right, leading to a separation of the predicted peaks larger than $0.5$. Thus the large phase shift can still be attributed to an inversion of the peaks.} The fit is not accurate, with a p-value of $0.01$. This might come from the second peak that is too thin compared to what is observed, but the p-value is still close from the threshold value and would be valid considering a criteria at $1\%$ confidence level. The modelled radio profile however does not recover the weak interpulse, probably for the same reason as for other pulsars in the previous sample (PSR~J1012$-$4235, PSR~J1125$-$6014 and PSR~J1909$-$3744). When a second fit on the interval given by \cite{shamohammadi_searches_2023} is done, the results are $\rchi_i=28.4 [+1.6,-1.4]\degr$, $i=74.5 [+0.5,-1.1]\degr$, and $\phi_i=0.495$. The value of the inclination found is consistent with the best fit $\zeta$ angle. Indeed we recover a very similar result for the $\gamma$-ray and radio profiles, and as for the previous fit the p-value is too small compared to our criteria, with a value of $0.002$.

\subsubsection{\cor{J0610-2100}} 

This pulsar of period $P=\SI{3.86}{\milli\second}$ \citep{van_der_wateren_irradiated_2022} shows two wide $\gamma$-ray peaks, and one radio peak with a weak interpulse. The best fit parameters are $\rchi=43.5 [+14.5,-18.3]\degr$, $\zeta=51.3 [+17.4,-13.3]\degr$, and $\phi=-0.25$. Considering the uncertainties in the data, the modelled $\gamma$-ray light curve fits rather well the second peak, but misses completely the first peak. \cortwo{Due to those large uncertainties in the data, we can not draw any firm conclusion concerning the large phase shift.} However, according to the KS test it is still accurate, with a p-value of $0.22$. Another solution that would be able to model both pulses is also possible, but it is not the one favoured by our method. Concerning the radio profile, the main pulse is rather well fitted, but we do not recover the interpulse, again for the same reason as given previously. The fitting using the constraint on the inclination from \cite{blanchard_census_2025} yields a similar fitting for the $\gamma$-ray curve, with $\rchi_i=34.3 [+7.7,-9.3]\degr$, $i=60.6 [+8.8,-6.6]\degr$, and $\phi_i=0.27$. The p-value is also of $0.22$. The light curve is slightly shifted to the right compared to previous fit, and the main radio pulse is wider. The best fit angles $i$ and $\zeta$ are compatible, so we conclude that there is indeed alignment for this pulsar.

\subsubsection{\cor{J0636+5128}} 

This pulsar of period $P=\SI{2.87}{\milli\second}$\citep{kaplan_dense_2018} shows two $\gamma$-ray peaks and one strong radio pulse. There is not enough date points for this $\gamma$-ray light curve to perform a KS test, but by eye we can already see that the fit does not recover the first $\gamma$-ray peak. Indeed the error bars are not restrictive enough to be able to detect the first peak with our method. Concerning the radio pulse profile, the pulse seems accurately recovered. The results of the fit are $\rchi=43.3 [+45.0,-25.7]\degr$, $\zeta=47.9 [+33.6,-28.8]\degr$, and $\phi = 0.46$. \cortwo{As for PSR J0610$-$2100, the uncertainties are too large to draw any conclusion on the large phase shift.} The fitting imposing the constraint on the inclination from \cite{blanchard_census_2025} yields a very similar result, with $\rchi_i=59.9 [+15.0,-7.3]\degr$, $i=25.7 [+2.2,-4.7]\degr$, and $\phi_i=0.46$. Due to too few points in the $\gamma$-ray light curve and the large error bars associated to them, the fitting does not add real constraints on top of the geometrical constraints imposed for the observation in the $\gamma$-ray and radio domains, and thus the best fit $\zeta$ is compatible with the value of the inclination found.

\subsubsection{\cor{J1124-3653}}
This pulsar of period $P=\SI{2.41}{\milli\second}$ shows two $\gamma$-ray peaks almost unresolved, and one wide radio peak with a large interpulse. The best fit parameters obtained are $\rchi=53.5 [+4.2,-4.4]\degr$, $\zeta=36.6 [+7.3,-3.6]\degr$, and $\phi=-0.025$. The fitting of the $\gamma$-ray light curve is accurate, with a p-value of $0.13$. The fit shows two unresolved $\gamma$-ray peaks that fit rather well the first peak, but we miss the second peak. The modelled radio pulse profile is rather good considering the noise, but we miss the interpulse. When the constraint on the inclination from \cite{blanchard_census_2025} is imposed, the result is similar, with $\rchi_i=50.4 [+0.6,-2.3]\degr$, $i=44.5 [+1.2,-0.5]\degr$, and $\phi_i=-0.015$. The p-value is now of $0.09$. No second $\gamma$-ray peak is predicted, as well as no radio interpulse, as for the previous fit. The inclination angle found is compatible with the hypothesis of alignment.

\subsubsection{\cor{J1514-4946}}
This pulsar of period $P=\SI{3.59}{\milli\second}$ shows close double $\gamma$-ray peaks and one radio peak. For this pulsar, the symmetrical solution has been kept. The fitting of the curves provides $\rchi= 27.8 [+0.2,-0.5]\degr$, $\zeta= 74.2 [+0.5,-0.2]\degr$, and $\phi= 0.015$. The fit looks rather accurate, however according to the KS test the p-value is only of $0.0006$. This could come from the amplitude of the modelled peak that is too small compared to what is observed. The radio pulse obtained is slightly too large. When a second fit imposing the constraint on the inclination from \cite{shamohammadi_searches_2023} is done, the $\gamma$-ray light curve obtained corresponds actually exactly to the previous one, with $\rchi_i=27.8 [+0.2,-0.5]\degr$, $i=74.2 [+0.5,-0.2]\degr$, and $\phi_i=0.015$, as well as the same p-value of $0.0006$. The radio profile is also the same as before. Thus this pulsar respects the hypothesis of alignment.

\subsubsection{\cor{J1544+4937}}
This pulsar of period $P=\SI{2.16}{\milli\second}$ \citep{mata_sanchez_black_2023} shows one wide $\gamma$-ray peak spread out over almost the whole period, and one radio peak. There are too few points to perform the KS test, but by eye we can see that the $\gamma$-ray fit seems accurate. The predicted radio profile is also acceptable. The best fit parameters are $\rchi=60.5 [+6.4,-25.6]\degr$, $\zeta=25.5 [+24.0,-9.5]\degr$, and $\phi=-0.02$. The fitting using the constraint given by \cite{blanchard_census_2025} yields a similar fitting for the $\gamma$-ray light curve, with a peak that is thinner but of similar amplitude. The radio pulse is also thinner than previously. The best fit parameters are $\rchi_i=44.1 [+9.0,-17.4]\degr$, $i=48.3 [+5.6,-5.3]\degr$, and $\phi_i=0.02$. Considering the large error bars on the best fit $\zeta$, the inclination found is still compatible with alignment.

\subsubsection{\cor{J1555-2908}}
This pulsar of period $P=\SI{1.79}{\milli\second}$ \citep{clark_neutron_2023} shows two $\gamma$-ray peaks, a main peak with a large amplitude and a second smaller one, as well as one large radio peak with an interpulse. The best fit parameters are $\rchi=67.4 [+4.2,-27.3]\degr$, $\zeta=70.7 [+12.7,-3.7]\degr$, and $\phi=0.37$. \cortwo{The separation of the two peaks being $\Delta=0.554\pm0.007$, the large phase shift can be attributed to an inversion of the peaks.} The fitting of the $\gamma$-ray light curve is accurate, with a p-value of $0.82$. The modelled radio profile recovers well the main radio pulse but does not predict the interpulse, probably due to the limitation of our model. The fitting imposing the constraint on the inclination from \cite{blanchard_census_2025} yields a very similar fit both in radio and in $\gamma$-ray, with $\rchi_i=58.5 [+6.4,-22.4]\degr$, $i=76.8 [+7.9,-1.8]\degr$, and $\phi_i=0.37$, which is compatible with the hypothesis of alignment. The p-value is of $0.82$ too.

\subsubsection{\cor{J1628-3205}}

This pulsar of period $P=\SI{3.21}{\milli\second}$ \citep{li_optical_2014, clark_neutron_2023} shows two wide $\gamma$-ray peaks spread out on the whole period, and one large radio pulse. The modelled $\gamma$-ray curve accurately fits the observed one, even though the amplitude of the first modelled peak is weak. The modelling of the radio pulse profile is rather good, even if the pulse is too wide. The best fit parameters are $\rchi=70.1 [+0.8,-2.1]\degr$, $\zeta=22.7 [+1.2,-2.1]\degr$, and $\phi=-0.01$, with a p-value of $0.43$,. The fit imposing the constraint on the inclination from \cite{blanchard_census_2025} yields an accurate $\gamma$-ray fit, however only the first $\gamma$-ray peak is modelled. The predicted radio pulse is now thinner than previously. The results for this fit are $\rchi_i=29.3 [+9.7,-9.3]\degr$, $i=62.6 [+7.4,-7.5]\degr$ and $\phi_i=-0.15$, with a p-value of $0.19$. The inclination is not compatible with the best fit $\zeta$ angle, meaning that this pulsar does not respect the hypothesis of alignment. We tried to look at the symmetrical solution, which gave a very similar result to the one obtained imposing the Shapiro constraint, with one $\gamma$-ray pulse missing. Since having two $\gamma$-ray peaks is physically more accurate, this observation favours the first fit without the Shapiro constraint.

\subsubsection{\cor{J1640+2224}}
This pulsar of period $P=\SI{3.16}{\milli\second}$ shows close double $\gamma$-ray peaks and one radio peak. The results of the fitting are $\rchi= 62.1 [+2.9,-30.8]\degr$, $\zeta= 34.7 [+30.1,-6.1]\degr$, and $\phi= -0.03$. The fit is accurate, with a p-value of $0.82$, even though the amplitude of the second $\gamma$-ray peak is not well recovered. The fit imposing the constrain of the Shapiro measurements \citep{lohmer_shapiro_2005, vigeland_bayesian_2014, fonseca_nanograv_2016} yields $\rchi_i=79.4 [+1.4,-5.2]\degr$, $i=71.0[+16.9,-17.0]\degr$, and $\phi_i=-0.17$. However looking at the distribution obtained for the inclination, we notice that our method gives no strong constraint on the possible value for the inclination for the interval given in the literature. For this fit, we obtain two well resolved peaks, which as seen by eye, does not correspond at all to the observed light curve. But due to the large error bars in the Fermi data, the fit is still considered as accurate, with a p-value of $0.16$. The modelled radio profile is also less accurate than previously, the main peak being slightly too thin and a small interpulse that is not observed is predicted. Considering the error bars on the value of the best fit $\zeta$ angle, this pulsar is aligned according to the measurements of \cite{fonseca_nanograv_2016} ($i=60\pm6\degr$), but not aligned according to the results of \cite{lohmer_shapiro_2005, vigeland_bayesian_2014} ($i=84[+4,-6]\degr$). Thus our solution favours the result of \cite{fonseca_nanograv_2016}, but we can not reach a definitive conclusion on whether or not this pulsar is aligned.

\subsubsection{\cor{J1732-5049}}
This pulsar of period $P=\SI{5.31}{\milli\second}$ shows one wide $\gamma$-ray peak spread over a significant fraction of the period, and one radio peak. The results of the fitting are $\rchi= 33.1 [+19.0,-6.7]\degr$, $\zeta= 62.9 [+6.1,-17.6]\degr$, and $\phi= 0.05$. The fitted light curve shows two unresolved $\gamma$-ray peaks rather than one large peak, but the fitting is still considered as accurate with a p-value of $0.53$. When the second fit imposing the constrain on the inclination from \cite{shamohammadi_searches_2023} is done, we obtain a very similar results with $\rchi_i = 32.4 [+5.4,-6.0]\degr$, $i=63.5 [+5.3,-4.5]\degr$, and $\phi_i=0.05$, compatible with the hypothesis of alignment. The p-value is the same as before, $0.53$. The modelled radio profile is very similar in both cases too, it does not reproduce the complex shape of the radio pulse but this is not what we are looking at.

\subsubsection{\cor{J1811-2405}}
This pulsar of period $P=\SI{2.7}{\milli\second}$ shows one weak and one stronger $\gamma$-ray peak, and a radio peak with a weak interpulse. The results of the fitting are $\rchi= 34.7 [+21,-6.7]\degr$, $\zeta= 77.8 [+3.2,-10.4]\degr$, and $\phi=-0.1$. The fit is accurate, with a p-value of $0.19$. The amplitude of the second peak is rather small compared to what is observed, but the first peak is well recovered. The predicted radio profile seems also accurate, we recover the interpulse even though its amplitude is small. The results of the fit imposing using Shapiro delay from \cite{ng_shapiro_2020} are $\rchi_i=31.5[+5.4,-6.9]\degr$, $i=75.6[+3.4,-2.6]\degr$, and $\phi_i=-0.07$. This is consistent with the best fit $\zeta$ angle obtained. We obtain a very similar result both in radio and in $\gamma$, only the first $\gamma$-ray peak is shifted to the right compared to what has been found previously. The corresponding p-value is $0.43$.

\subsubsection{\cor{J1959+2048}}
This pulsar of period $P=\SI{1.61}{\milli\second}$ shows one main thin $\gamma$-ray peak, as well as a second wider $\gamma$-ray pulse with a smaller amplitude. It also presents a complex radio pulse profile with three pulses : two strong ones separated by slightly less than half a period, and one small interpulse close to one of the strong pulse. The results of the fit are $\rchi=32.1 [+15.7,-4.2]\degr$, $\zeta=85.2 [+1.7,-4.9]\degr$, and $\phi=0.325$. \cortwo{The separation of the two peaks being $\Delta=0.566\pm0.007$, the large phase shift can be attributed to an inversion of the peaks. Another interpretation would be to identify the second radio pulse as the main pulse for phase-aligning the gamma rays. In this case, the radio peak at phase $\varphi\approx0.3$ would genuinely be the one at phase $\varphi\approx0$.} The fit is accurate, with a p-value of $0.41$. The second $\gamma$-ray peak is rather thin compared to the observed one, but it is well centred on it. The predicted radio profile is rather accurate : the main radio peak is a wide but well centred, however the interpulse has a complex shape composed of several peaks that are not all recovered, since only one interpulse can be predicted by the model. The fit using Shapiro delay constraints from \cite{blanchard_census_2025} deviates somehow from the previous fit, with $\rchi_i=69.5 [+2.4,-0.5]\degr$, $i=66.5 [+0.5,-0.8]\degr$, and $\phi_i=0.325$, with a p-value of $0.16$. The $ \gamma$-ray light curve obtained is similar to the previous one, only with peaks having a larger amplitude, but the inclination obtained is not compatible with the best fit $\zeta$ angle. The only strong difference between the two fits is that in the case where $\zeta$ is left as a free parameter, the predicted radio profile is able to model the interpulse. This is not the case for the second fit using the Shapiro delay constraint. Using this argument, we could favour the first fit and conclude that this pulsar does not respect the hypothesis of alignment. However, considering the article of \cite{kramer_radio_2025}, the situation might be more complex.

Our model for the prediction of the radio pulse profile assumes that the emission originates from the polar caps. However according to \cite{kramer_radio_2025}, radio emission could also occur at the same place as $\gamma$ emission, which would result in radio and $\gamma$-ray pulses aligned in phase. For PSR J1959$+$2048, the article concludes that both the narrow pulse and the weaker and wide pulse aligned in phase with the $\gamma$ emission might originate from the light cylinder, whereas only the third narrow pulse would come from the polar caps. In that case since our model only predicts polar cap emission, it would not be able to appropriately predict the radio emission for this pulsar, and would not consist of a strong enough argument to exclude the hypothesis of alignment.

\subsubsection{\cor{J2051-0827}}
This pulsar of period $P=\SI{4.51}{\milli\second}$ shows one wide $\gamma$-ray peak spread out on the whole period, and one radio peak with two components. As for PSR~J0101$-$6422 and PSR~J1514$-$4946, the symmetrical solution has been taken. The best fit parameters are $\rchi=26.8 [+3.3,-1.3]\degr$, $\zeta=69.2 [+1.2,-3.5]\degr$, and $\phi=0.05$. The fit of the $\gamma$-ray light curve is accurate, with a p-value of $0.43$, however the predicted radio pulse overestimates slightly the signal between the pulses and the pulse is too wide. The fitting imposing the constraint from \cite{blanchard_census_2025} on the inclination is still accurate, with $\rchi_i=25.2 [+3.8,-5.3]\degr$, $i=55.5 [+5.4,-4.2]\degr$, and $\phi_i=0.05$, with a p-value of $0.16$. In this case the $\gamma$-ray pulse is too thin compared to the observed one, but the radio profile predicted is more precise. Finally the best fit $\zeta$ angle is only $5\degr$ away from the best fit $i$, considering the error bars.

\subsubsection{\cor{J2256-1024}}
This pulsar of period $P=\SI{2.29}{\milli\second}$ shows two $\gamma$-ray peaks : one strong and thin pulse, and a very small one. It also presents a complex radio peak with three components, and a very small interpulse. The symmetrical solution has been kept for this pulsar. The best fit parameters are $\rchi=37.1 [+7.9,-1.37]\degr$, $\zeta=88.4 [+1.4,-3.6]\degr$, and $\phi=0.05$. The fitting of the $\gamma$-ray light curve is accurate, with a p-value of $0.1$, even though the amplitude of the first $\gamma$-ray peak that is too small compared to observations. Concerning the predicted radio pulse profile, the main peak is rather well reconstructed even if the substructures of the pulse are not recovered, an interpulse is predicted but with an amplitude too large compared to the observed one. The fit using Shapiro delay constraints from \cite{blanchard_census_2025} gives $\rchi_i=84.4 [+4.5,-1.4]\degr$, $i=64.8 [+2.7,-2.8]\degr$, and $\phi_i=-0.04$, with a p-value of $0.05$. The fit is similar to the previous one, but the amplitude of the first $\gamma$-ray peak is closer to observations. However, the second peak is thin compared to the observed one. The modelled radio profile follows more accurately the observed one, with a thinner main pulse and an interpulse smaller in amplitude. However the possible interval found for the inclination is not in agreement with the one found for the $\zeta$ angle, thus this pulsar does not respect the hypothesis of alignment.

\subsection{Discussion}

\begin{figure*}[h]
    \centering
    \includegraphics[width=\linewidth]{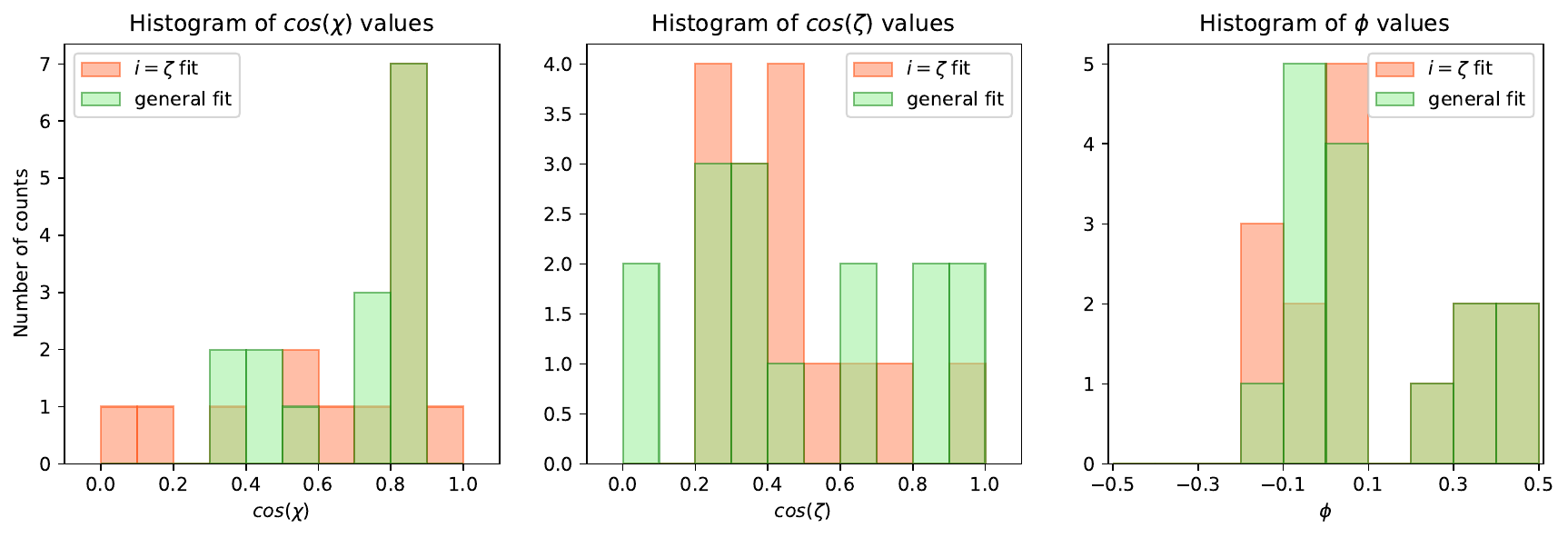}
    \caption{Same as Fig.~\ref{fig:MSP1-histogram} but for the second sample of MSPs.}
    \label{fig:MSP2-histogram}
\end{figure*}

\begin{figure*}[h]
    \centering
    \includegraphics[width=\linewidth]{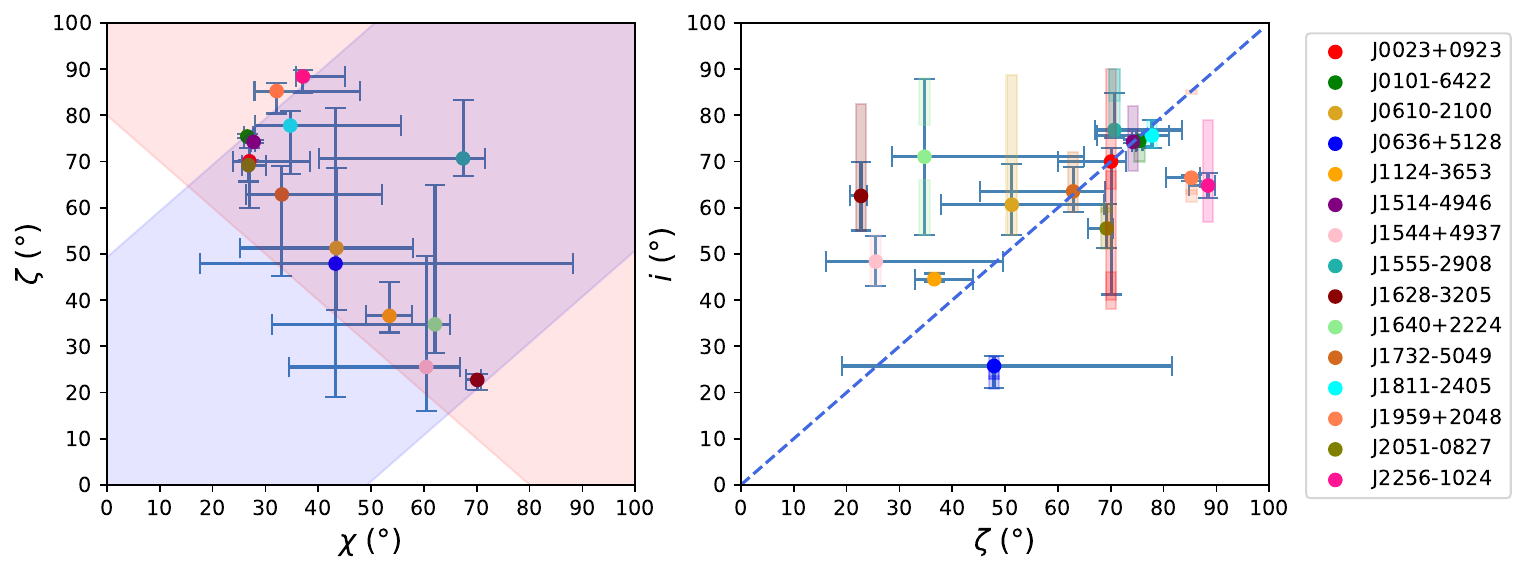}
    \caption{\cor{Same as Fig.~\ref{fig:MSP2-chi-i-zeta-plane}} but for the second sample of MSPs.}
    \label{fig:MSP2-chi-i-zeta-plane}
\end{figure*}

Histograms shown in figure~\ref{fig:MSP2-histogram} summarise the best fit parameters, on the left for~\cor{$\cos\rchi$}, in the middle for~\cor{$\cos\zeta$} and on the right for~$\phi$, in green in the unconstrained case and in orange when the condition $i=\zeta$ is imposed. Finally, figure~\ref{fig:MSP2-chi-i-zeta-plane} shows the best fit parameters found for this sample. On the left is a $\rchi-\zeta$ plane showing the best fitting couple of angles $\rchi$ and $\zeta$ for each pulsar, and on the right is a $\zeta-i$ plane showing the best fit for those two angles for each pulsar. The \cortwo{coloured} error bars correspond to the different inclination angle measurements that are available in the literature (table \ref{tab:Sample_Shapiro2}), and the dark blue error bars to the uncertainty obtained after the fitting of the light curves. \cor{The uncertainty on the inclination~$i$ has been reduced with the use of our method for most of the pulsars from the sample.} For this sample, \cor{$87\%$} of the MSPs show a spin-orbit alignment, \cor{including PSR~J2051$-$0827 that is still $5\degr$ away from alignment. Excluding this pulsar would lower to $80\%$ the percentage of alignment. There are two pulsars for which we have reasonable clues to expect strong misalignment : PSR J1628$-$3205 and PSR~J2256$-$1024.} There are several possible reasons to expect misalignment, as we finally discuss in the next section. 

\section{\cortwo{Discussion about large phase shift\label{sec:discussion-non-radial-prop}}}

\begin{figure*}[h]
    \centering
    \includegraphics[width=0.7\linewidth]{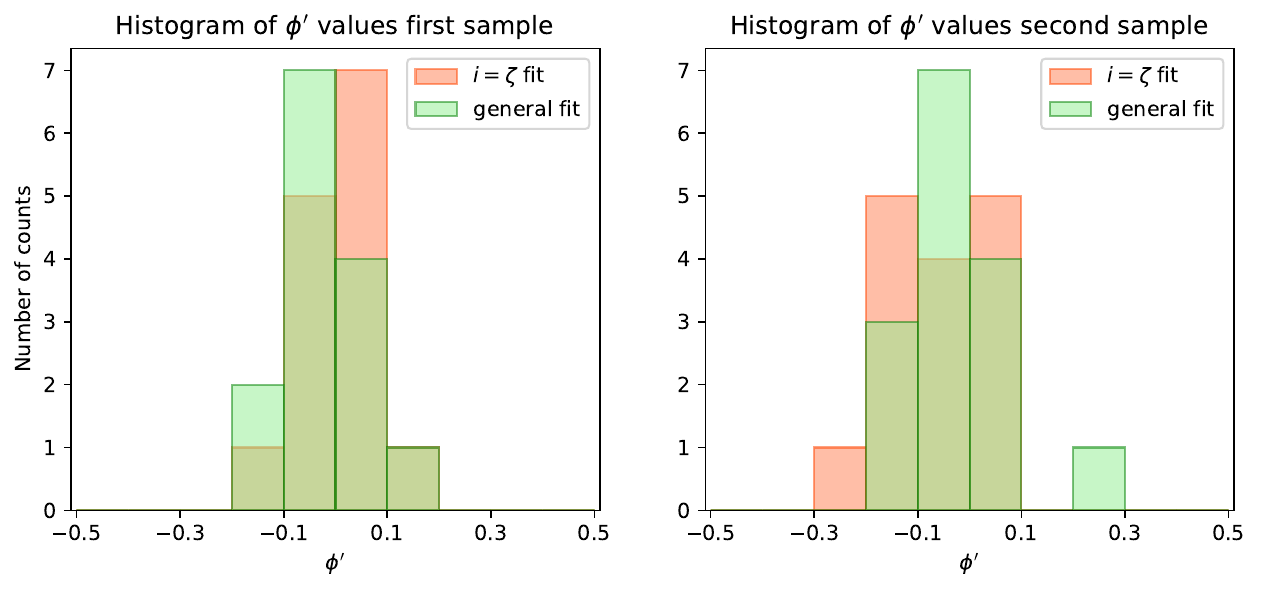}
    \caption{\cortwo{Histogram of best fit corrected phase shifts $\phi'$ for the first pulsar sample (on the left) and the second pulsar sample (on the right).}}
    \label{fig:compl-angle-histogram}
\end{figure*}

\cortwo{As already stated, the parameter~$\phi$ is mainly used to correct for a possible non-radial propagation effect not included in our emission model. Indeed, the force-free/striped wind model should not be confused with the associated emission model because it only gives the geometry of the current sheet, i.e. the region where plasma significantly radiates. However, it does not prescribe the direction into which these particles are emitting. An azimuthal velocity component could lead to a relativistic beaming direction varying with the distance to the light cylinder and as a consequence a variation in the $\gamma$-ray light curve profiles to some extent. However non-radial propagation can only explain phase shifts up to $\pm0.25$. Indeed the phase shift depends linearly on the angle between the radial direction and the particle propagation direction. This angle, written as $\Delta\theta$, is expressed as \citep{petri_multi-wavelength_2024} : $\rm tan(\Delta\theta) = {v_\varphi}/{v_r}$, with $v_r$ and $v_\varphi$ being respectively the velocity of the emitted particles projected along the radial and azimuthal axis. The phase shift $\phi$ is then linked to $\Delta\theta$ by the relation : $\phi={\Delta\theta}/{2\pi}$. To give some orders of magnitudes in degrees, a phase shift of $0.1$ would need a deviation from the radial propagation of $\Delta\theta=36\degr$, a phase shift of $0.2$ would need a deviation of $\Delta\theta=72\degr$, and so on. For more details on the impact of a non-radial flow on the radio time lag, see the discussion in Section~5.1 of \cite{petri_multi-wavelength_2024}, and in particular Fig.11 in that paper.}

\cortwo{For pulsars having phase shift outside the interval of $[-0.25,0.25]$, we interpret this as having an inversion of the peaks due to the degeneracy in two-pulse $\gamma$-ray light curves. This means that to define properly the phase zero of profiles showing a radio interpulse, it would be more appropriate to choose the radio pulse implying a separation of the $\gamma$-ray peaks smaller than half a period, rather than systematically taking the main radio pulse. The inversion of the peaks requires a shift of half the period, thus to only evaluate the effect of non-radial propagation for those pulsars, we take the complementary of their best fit $\phi$ on the interval of half the period to which they belong ($[0,0.5]$ or $[-0.5,0]$) : for a positive $\phi$, the complementary $\phi'$ is given by $\phi'=\phi-0.5$, and for a negative $\phi$ the complementary $\phi'$ is given by : $\phi'=\phi+0.5$. The histograms of $\phi'$ values obtained for both samples are shown on figure \ref{fig:compl-angle-histogram}. For the first sample, the distribution peaks at a mean value of $-0.01$ and has median of $-0.01$. For the second sample, the mean of the distribution is $-0.02$ and the median equals to $-0.02$, meaning that in general the pulses arrive at a phase as predicted by the model. Thus in general, assuming a radial propagation for the pulsar wind is a good first approximation to the real flow.}

\section{Discussion about possible misalignment\label{sec:Discussion}}

\cor{Spin-orbit alignment arises due to the accretion of matter and therefore angular momentum from an accretion disk whose angular momentum vector is orthogonal to the orbital plane that overlaps with the accretion disk plane. The accreted angular momentum $\Delta L$ must be comparable to the pulsar spin angular momentum~$L$ in order to expect significant evolution towards alignment, thus $\Delta L \approx L = I\,\Omega$, with $I=\frac{2}{5}M R^2$ the moment of inertia of the pulsar ($M$ and $R$ being respectively the mass and radius of the pulsar), and $\Omega$ the angular velocity. \cor{We assumed a homogeneous density inside the star for the computation of the moment of inertia}. The evolution of the pulsar angular momentum is then given by
\begin{equation}
\label{eq:theoreme_moment_cinetique}
    \frac{dL}{dt} = I\, \frac{d\Omega}{dt}+\frac{dI\,}{dt}\Omega = N
\end{equation}
with $N$ the torque acting on the pulsar due to the accretion of matter from the companion. A simple estimate assuming a constant accretion rate of $\dot{m}$ at a radius $r_{\rm acc}$ shows that $N=\dot m \, \sqrt{G\, M \, r_{\rm acc}}$ with the internal limit of the accretion disk designates as the \cor{accretion} radius $r_{\rm acc}$. Moreover, as explained in details in \cite{abolmasov_spin_2024}, in order for the pulsar to be in an accretion regime, this radius must be smaller than the corotation radius $R_{\rm co}$ and \cor{than} the gravitational capture radius $R_G$. }

Taking a pulsar of initial mass $M_0$ and increasing with time as $M(t) = M_0 + \dot m \, t$, Eq.~\eqref{eq:theoreme_moment_cinetique} is rewritten as
\begin{equation}
    \frac{dL}{dt} = I\, \dot \Omega + I\, \frac{\dot m}{M} \Omega = \dot m \, \sqrt{G\,M \, r_{\rm acc}}
\end{equation}
using the fact that ${\dot I}/{I}={\dot m}/{M}$. Its integration yields the general expression for the stellar angular velocity as
\begin{equation}
    \Omega(t) = \frac{\Omega_0}{1+ {\dot m\,t}/{M_0}} + \frac{5}{3R^2}\sqrt{G \, M_0 \, r_{\rm acc}} \left( \sqrt{1+\frac{\dot m}{M_0} \, t}-\frac{1}{1+{\dot m\,t}/{M_0}} \right)
\end{equation}
with $\Omega_0$ corresponding to the initial angular velocity of the pulsar, taken to be $\Omega_0=\frac{2 \pi}{P_0}$ with $P_0$ its initial rotation period. The associated general expression for the angular momentum is
\begin{equation}
    L(t) = I(t) \, \Omega(t) = L_0 + \frac{2}{3}\sqrt{G \, M_0^3 \, r_{\rm acc}} \left[ \left(1+\frac{\dot m}{M_0} \, t \right)^{3/2}-1\right]
\end{equation}
with $L_0 = I_0 \, \Omega_0 = \frac{2}{5} M_0 \, R^2 \,\Omega_0$ the initial angular momentum. The typical alignment time scale $\tau_{\rm align}$ for which we would have a variation of the angular momentum of the order $\Delta L \approx L_0$ can then be estimated by
\begin{equation}
    \frac{\Delta L}{L_0} = \frac{5 \, P_0 \, \dot m }{4 \, \pi \, R^2} \, \tau_{\rm align} \, \sqrt{\frac{G\,r_{\rm acc}}{M_0} \left(1+\frac{\dot m}{M_0} \, \tau_{\rm align}\right)} \approx 1 \ .
\end{equation}
Assuming a spherical accretion, we take for the inner radius of the accretion disk the expression (see \cite{biryukov_magnetic_2021})
\begin{equation}
 r_{\rm acc} = \frac{1}{2} \left( \frac{\mu^4}{2\,G \, M_0 \, \dot m^2} \right)^{1/7}     
\end{equation}
where $\mu$ is the stellar magnetic moment. The latest can be estimated with the formula $\mu^2 = 4\pi \frac{B^2 R^6}{\mu_0}$, where $B$ is the strength of the dipolar magnetic field \cor{at the equator}. This way the previous equation can be recast into
\begin{equation}
\label{eq:alignement}
\begin{aligned}
    \tau_{\rm align}= \SI{3.8e6}{\year} \left(\frac{P_0}{\SI{1}{\milli\second}}\right)^{-1} \left(\frac{R}{\SI{10}{\kilo\meter}}\right)^{8/7} \left( \frac{M_0}{1.4 M_{\odot}} \right)^{4/7} \\ \left(\frac{\dot m}{\dot M_{\rm Edd}}\right)^{-6/7} \left( \frac{B}{\SI{e6}{\tesla}} \right)^{-2/7}
\end{aligned}
\end{equation}
with $\dot M_{\rm Edd}$ the Eddington accretion rate. To obtain this form, we assumed that $\frac{\dot m \,\tau_{\rm align}}{M_0} \ll 1$. Indeed, in a first approximation we can write $\dot m \approx \frac{\Delta M}{\tau_{\rm align}}$, with $\Delta M$ the variation of mass of the pulsar, meaning that $\frac{\dot m \, \tau_{\rm align}}{M_0} \approx \frac{\Delta M}{M_0}$. As explained in \cite{tauris_formation_2017}, the total mass accreted by the pulsar during \cor{its} accretion phase represents only a small fraction of \cor{its} total mass. \cor{Thus} we can assume $\frac{\Delta M}{M_0} \ll 1$. \cor{The expression~\eqref{eq:alignement} gives} an estimate of the typical time needed to reach alignment, which is of the order of \SI{3.8e6}{\year} \cor{for typical pulsar binary parameters}.

In most cases, when the accretion phase starts, the period of the pulsar will be much larger than $P=\SI{1}{\milli\second}$ as given in the former formula, meaning that the previous result rather corresponds to an upper limit of this alignment time. By taking a more realistic value for the initial period, that is $P=\SI{5}{\second}$, we now obtain a new value for the alignment time of the order of \SI{1}{\kilo\year}. An accretion rate well below the Eddington limit could also drastically increase the alignment time scale by a factor 10 to 100.

\begin{figure}[h]
    \centering
    \includegraphics[width=0.9\linewidth]{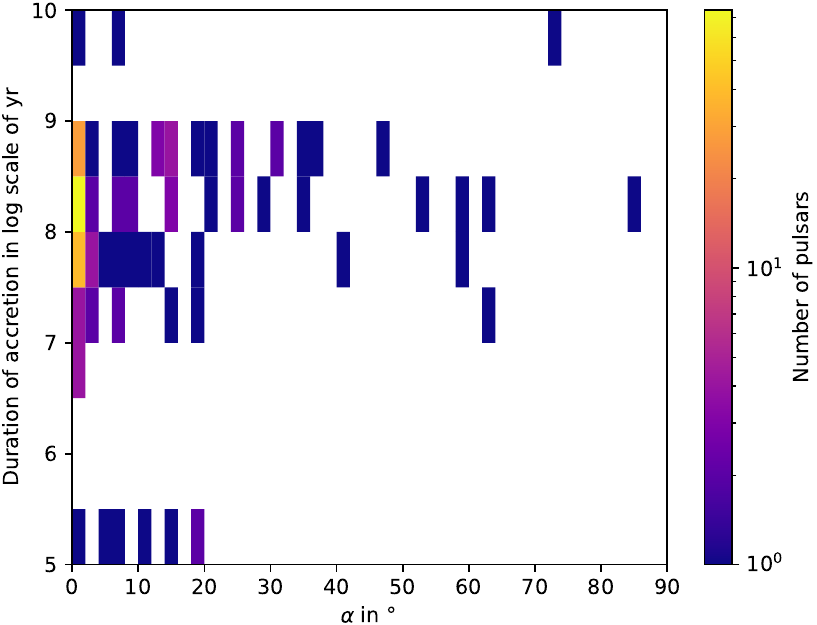}
    \caption{Spin-orbit inclination angle $\alpha$ of a simulated population of MSPs that are still in a binary after the end of the accretion phase. The majority of MSPs show close to alignment geometries after tens of millions of years.}
    \label{fig:2D_map_tacc_beta}
\end{figure}

To assess whether a pulsar had enough time to reach alignment during the accretion phase, we need an estimate of the total duration of that phase. For that purpose, we simulated a population of millisecond pulsars (MSPs) born in binaries with main-sequence companions, modelling their evolution using the SEVN code \citep{spera_very_2017,iorio_compact_2023} to track both stellar and binary processes. The spin and magnetic evolution of pulsars was treated following the prescription of \citet{biryukov_magnetic_2021}. To identify detectable systems, we applied a detection pipeline closely following \citet{sautron_galactic_2024}, accounting for selection effects in both radio (at \SI{1.4}{\giga\hertz}) and $\gamma$-ray bands, based on Fermi/LAT sensitivity. Full details of the simulation and detection methodology are presented in \cite{sautron_born_2025}. One of the outcomes of the simulation, which involved 580\,000 binaries and led to 519 detections (either in radio or $\gamma$-ray, or both) that matched the observations in the $P-\dot{P}$ diagram, with only 240 pulsars remaining in binaries, is \cor{plotted} in figure~\ref{fig:2D_map_tacc_beta}. This figure shows the number of pulsars \cor{left} at the end of the simulation depending on the angle~$\alpha$ between the rotation axis and the orbital angular momentum of the binary, indicated on the horizontal axis, and on the duration of the accretion phase, indicated on the vertical axis. For the large majority of pulsars, the duration of the accretion phase is about $10^{8-9}$~yr, \cor{thus} much larger than the orders of magnitude we found for the alignment time in the previous calculation. For this reason, we expect most of the pulsars to reach alignment during the accretion phase. Figure~\ref{fig:2D_map_tacc_beta} also confirms that point, since \cor{almost all pulsars ended with an angle $\alpha$ situated between $0\degr$ and $10\degr$, after an accretion regime of duration $10^{8-9}$~yr}. However, 20~\% of the simulated pulsars finished their accretion phase with a residual angle $\alpha$ superior to $10\degr$. If we exclude the pulsars for which the misalignment we found most probably comes from some limitations of our $\gamma$-ray light curve model, it seems indeed possible that several MSPs are misaligned in our sample. 


\section{Conclusion\label{sec:Conclusion}}

MSPs are recycled pulsars that are spun up by accretion of matter from their companion in a binary system. This spin up is accompanied by a torque tending to align the stellar spin with the orbital angular momentum vector on a rather short time scale. Being old pulsars with ages of around a billion year, they had enough time to align their spin with the orbit. This hypothesis has been check in our work by computing the pulsar viewing angle from $\gamma$-ray light curve fitting. By using two samples, one with accurate measurements of the orbital inclination angle~$i$ and another with some constraints on the possible intervals, we found that \cor{about} 4/5 of the MSPs indeed show spin-orbit alignment within several degrees. However a small subsample cannot be accommodated with this hypothesis, thus either our $\gamma$-ray light curve modelling is inaccurate or they indeed are largely misaligned. This possible misalignment has been found in our MSP population synthesis simulation where a small fraction of pulsars still have a discrepancy greater than $10\degr$ between the rotation axis and normal to the orbital plane. The origin of this misalignment must be accounted for, by for instance the accretion history and by the initial condition before the accretion regime. However such detailed analysis of the binary evolution and outcome will be the focus of a forthcoming investigation based on MSP population synthesis.

\begin{acknowledgements}
This work has been supported by the French Research Agency grant ANR-20-CE31-0010. We are grateful to our referee for his/her helpful comments. 
\end{acknowledgements}

\bibliographystyle{bibtex/aa}
\bibliography{Biblio-article}

@article{benli_constraining_2021,
	title = {Constraining millisecond pulsar geometry using time-aligned radio and gamma-ray pulse profile},
	volume = {647},
	issn = {0004-6361},
	url = {https://ui.adsabs.harvard.edu/abs/2021A&A...647A.101B},
	doi = {10.1051/0004-6361/202039853},
	urldate = {2025-04-14},
	journal = {A\&A},
	author = {Benli, Onur and Pétri, Jérôme and Mitra, Dipanjan},
	month = mar,
	year = {2021},
	pages = {A101},
}

@article{guillemot_non-detection_2014,
	title = {On the non-detection of γ-rays from energetic millisecond pulsars - dependence on viewing geometry},
	volume = {439},
	issn = {0035-8711},
	url = {https://ui.adsabs.harvard.edu/abs/2014MNRAS.439.2033G},
	doi = {10.1093/mnras/stu082},
	urldate = {2025-04-23},
	journal = {MNRAS},
	author = {Guillemot, L. and Tauris, T. M.},
	month = apr,
	year = {2014},
	pages = {2033--2042},
}

@article{biryukov_magnetic_2021,
	title = {Magnetic angle evolution in accreting neutron stars},
	volume = {505},
	issn = {0035-8711},
	url = {https://ui.adsabs.harvard.edu/abs/2021MNRAS.505.1775B},
	doi = {10.1093/mnras/stab1378},
	urldate = {2025-04-23},
	journal = {MNRAS},
	author = {Biryukov, Anton and Abolmasov, Pavel},
	month = aug,
	year = {2021},
	pages = {1775--1786},
}

@article{smith_third_2023,
	title = {The {Third} {Fermi} {Large} {Area} {Telescope} {Catalog} of {Gamma}-{Ray} {Pulsars}},
	volume = {958},
	issn = {0004-637X},
	url = {https://ui.adsabs.harvard.edu/abs/2023ApJ...958..191S},
	doi = {10.3847/1538-4357/acee67},
	urldate = {2025-04-23},
	journal = {ApJ},
	author = {Smith, D. A. and Abdollahi, S. and Ajello, M. and Bailes, M. and Baldini, L. and Ballet, J. and Baring, M. G. and Bassa, C. and Gonzalez, J. Becerra and Bellazzini, R. and Berretta, A. and Bhattacharyya, B. and Bissaldi, E. and Bonino, R. and Bottacini, E. and Bregeon, J. and Bruel, P. and Burgay, M. and Burnett, T. H. and Cameron, R. A. and Camilo, F. and Caputo, R. and Caraveo, P. A. and Cavazzuti, E. and Chiaro, G. and Ciprini, S. and Clark, C. J. and Cognard, I. and Corongiu, A. and Orestano, P. Cristarella and Crnogorcevic, M. and Cuoco, A. and Cutini, S. and D'Ammando, F. and de Angelis, A. and DeCesar, M. E. and De Gaetano, S. and de Menezes, R. and Deneva, J. and de Palma, F. and Di Lalla, N. and Dirirsa, F. and Di Venere, L. and Domínguez, A. and Dumora, D. and Fegan, S. J. and Ferrara, E. C. and Fiori, A. and Fleischhack, H. and Flynn, C. and Franckowiak, A. and Freire, P. C. C. and Fukazawa, Y. and Fusco, P. and Galanti, G. and Gammaldi, V. and Gargano, F. and Gasparrini, D. and Giacchino, F. and Giglietto, N. and Giordano, F. and Giroletti, M. and Green, D. and Grenier, I. A. and Guillemot, L. and Guiriec, S. and Gustafsson, M. and Harding, A. K. and Hays, E. and Hewitt, J. W. and Horan, D. and Hou, X. and Jankowski, F. and Johnson, R. P. and Johnson, T. J. and Johnston, S. and Kataoka, J. and Keith, M. J. and Kerr, M. and Kramer, M. and Kuss, M. and Latronico, L. and Lee, S. -H. and Li, D. and Li, J. and Limyansky, B. and Longo, F. and Loparco, F. and Lorusso, L. and Lovellette, M. N. and Lower, M. and Lubrano, P. and Lyne, A. G. and Maan, Y. and Maldera, S. and Manchester, R. N. and Manfreda, A. and Marelli, M. and Martí-Devesa, G. and Mazziotta, M. N. and McEnery, J. E. and Mereu, I. and Michelson, P. F. and Mickaliger, M. and Mitthumsiri, W. and Mizuno, T. and Moiseev, A. A. and Monzani, M. E. and Morselli, A. and Negro, M. and Nemmen, R. and Nieder, L. and Nuss, E. and Omodei, N. and Orienti, M. and Orlando, E. and Ormes, J. F. and Palatiello, M. and Paneque, D. and Panzarini, G. and Parthasarathy, A. and Persic, M. and Pesce-Rollins, M. and Pillera, R. and Poon, H. and Porter, T. A. and Possenti, A. and Principe, G. and Rainò, S. and Rando, R. and Ransom, S. M. and Ray, P. S. and Razzano, M. and Razzaque, S. and Reimer, A. and Reimer, O. and Renault-Tinacci, N. and Romani, R. W. and Sánchez-Conde, M. and Parkinson, P. M. Saz and Scotton, L. and Serini, D. and Sgrò, C. and Shannon, R. and Sharma, V. and Shen, Z. and Siskind, E. J. and Spandre, G. and Spinelli, P. and Stappers, B. W. and Stephens, T. E. and Suson, D. J. and Tabassum, S. and Tajima, H. and Tak, D. and Theureau, G. and Thompson, D. J. and Tibolla, O. and Torres, D. F. and Valverde, J. and Venter, C. and Wadiasingh, Z. and Wang, N. and Wang, N. and Wang, P. and Weltevrede, P. and Wood, K. and Yan, J. and Zaharijas, G. and Zhang, C. and Zhu, W.},
	month = dec,
	year = {2023},
	pages = {191},
}

@article{shamohammadi_searches_2023,
	title = {Searches for {Shapiro} delay in seven binary pulsars using the {MeerKAT} telescope},
	volume = {520},
	issn = {0035-8711},
	url = {https://ui.adsabs.harvard.edu/abs/2023MNRAS.520.1789S},
	doi = {10.1093/mnras/stac3719},
	urldate = {2025-04-23},
	journal = {MNRAS},
	author = {Shamohammadi, M. and Bailes, M. and Freire, P. C. C. and Parthasarathy, A. and Reardon, D. J. and Shannon, R. M. and Venkatraman Krishnan, V. and Bernadich, M. C. i. and Cameron, A. D. and Champion, D. J. and Corongiu, A. and Flynn, C. and Geyer, M. and Kramer, M. and Miles, M. T. and Possenti, A. and Spiewak, R.},
	month = apr,
	year = {2023},
	pages = {1789--1806},
}

@article{tan_high-cadence_2024,
	title = {High-cadence {Timing} of {Binary} {Pulsars} with {CHIME}},
	volume = {966},
	issn = {0004-637X},
	url = {https://ui.adsabs.harvard.edu/abs/2024ApJ...966...26T},
	doi = {10.3847/1538-4357/ad28b2},
	urldate = {2025-04-23},
	journal = {ApJ},
	author = {Tan, Chia Min and Fonseca, Emmanuel and Crowter, Kathryn and Dong, Fengqiu Adam and Kaspi, Victoria M. and Masui, Kiyoshi W. and McKee, James W. and Meyers, Bradley W. and Ransom, Scott M. and Stairs, Ingrid H.},
	month = may,
	year = {2024},
	pages = {26},
}

@article{serylak_eccentric_2022,
	title = {The eccentric millisecond pulsar, {PSR} {J0955}−6150. {I}. {Pulse} profile analysis, mass measurements, and constraints on binary evolution},
	volume = {665},
	issn = {0004-6361},
	url = {https://ui.adsabs.harvard.edu/abs/2022A&A...665A..53S},
	doi = {10.1051/0004-6361/202142670},
	urldate = {2025-04-23},
	journal = {A\&A},
	author = {Serylak, M. and Venkatraman Krishnan, V. and Freire, P. C. C. and Tauris, T. M. and Kramer, M. and Geyer, M. and Parthasarathy, A. and Bailes, M. and Bernadich, M. C. i. and Buchner, S. and Burgay, M. and Camilo, F. and Karastergiou, A. and Lower, M. E. and Possenti, A. and Reardon, D. J. and Shannon, R. M. and Spiewak, R. and Stairs, I. H. and van Straten, W.},
	month = sep,
	year = {2022},
	pages = {A53},
}

@article{gautam_detection_2024,
	title = {Detection of the relativistic {Shapiro} delay in a highly inclined millisecond pulsar binary {PSR} {J1012}−4235},
	volume = {682},
	issn = {0004-6361},
	url = {https://ui.adsabs.harvard.edu/abs/2024A&A...682A.103G},
	doi = {10.1051/0004-6361/202347836},
	urldate = {2025-04-23},
	journal = {A\&A},
	author = {Gautam, T. and Freire, P. C. C. and Wu, J. and Venkatraman Krishnan, V. and Kramer, M. and Barr, E. D. and Bailes, M. and Cameron, A. D.},
	month = feb,
	year = {2024},
	pages = {A103},
}

@article{desvignes_high-precision_2016,
	title = {High-precision timing of 42 millisecond pulsars with the {European} {Pulsar} {Timing} {Array}},
	volume = {458},
	doi = {10.1093/mnras/stw483},
	number = {3},
	journal = {MNRAS},
	author = {Desvignes, G. and Caballero, R. N. and Lentati, L. and Verbiest, J. P. W. and Champion, D. J. and Stappers, B. W. and Janssen, G. H. and Lazarus, P. and Osłowski, S. and Babak, S. and Bassa, C. G. and Brem, P. and Burgay, M. and Cognard, I. and Gair, J. R. and Graikou, E. and Guillemot, L. and Hessels, J. W. T. and Jessner, A. and Jordan, C. and Karuppusamy, R. and Kramer, M. and Lassus, A. and Lazaridis, K. and Lee, K. J. and Liu, K. and Lyne, A. G. and McKee, J. and Mingarelli, C. M. F. and Perrodin, D. and Petiteau, A. and Possenti, A. and Purver, M. B. and Rosado, P. A. and Sanidas, S. and Sesana, A. and Shaifullah, G. and Smits, R. and Taylor, S. R. and Theureau, G. and Tiburzi, C. and van Haasteren, R. and Vecchio, A.},
	month = may,
	year = {2016},
	pages = {3341--3380},
}

@article{arzoumanian_nanograv_2018,
	title = {The {NANOGrav} 11 {Year} {Data} {Set}: {Pulsar}-timing {Constraints} on the {Stochastic} {Gravitational}-wave {Background}},
	volume = {859},
	issn = {0004-637X},
	shorttitle = {The {NANOGrav} 11 {Year} {Data} {Set}},
	url = {https://ui.adsabs.harvard.edu/abs/2018ApJ...859...47A},
	doi = {10.3847/1538-4357/aabd3b},
	urldate = {2025-04-23},
	journal = {ApJ},
	author = {Arzoumanian, Z. and Baker, P. T. and Brazier, A. and Burke-Spolaor, S. and Chamberlin, S. J. and Chatterjee, S. and Christy, B. and Cordes, J. M. and Cornish, N. J. and Crawford, F. and Thankful Cromartie, H. and Crowter, K. and DeCesar, M. and Demorest, P. B. and Dolch, T. and Ellis, J. A. and Ferdman, R. D. and Ferrara, E. and Folkner, W. M. and Fonseca, E. and Garver-Daniels, N. and Gentile, P. A. and Haas, R. and Hazboun, J. S. and Huerta, E. A. and Islo, K. and Jones, G. and Jones, M. L. and Kaplan, D. L. and Kaspi, V. M. and Lam, M. T. and Lazio, T. J. W. and Levin, L. and Lommen, A. N. and Lorimer, D. R. and Luo, J. and Lynch, R. S. and Madison, D. R. and McLaughlin, M. A. and McWilliams, S. T. and Mingarelli, C. M. F. and Ng, C. and Nice, D. J. and Park, R. S. and Pennucci, T. T. and Pol, N. S. and Ransom, S. M. and Ray, P. S. and Rasskazov, A. and Siemens, X. and Simon, J. and Spiewak, R. and Stairs, I. H. and Stinebring, D. R. and Stovall, K. and Swiggum, J. and Taylor, S. R. and Vallisneri, M. and van Haasteren, R. and Vigeland, S. and Zhu, W. W. and {NANOGrav Collaboration}},
	month = may,
	year = {2018},
	pages = {47},
}

@article{ng_shapiro_2020,
	title = {A {Shapiro} delay detection in the pulsar binary system {PSR} {J1811}-2405},
	volume = {493},
	issn = {0035-8711},
	url = {https://ui.adsabs.harvard.edu/abs/2020MNRAS.493.1261N},
	doi = {10.1093/mnras/staa337},
	urldate = {2025-04-23},
	journal = {MNRAS},
	author = {Ng, C. and Guillemot, L. and Freire, P. C. C. and Kramer, M. and Champion, D. J. and Cognard, I. and Theureau, G. and Barr, E. D.},
	month = mar,
	year = {2020},
	pages = {1261--1267},
}

@article{breton_discovery_2013,
	title = {Discovery of the {Optical} {Counterparts} to {Four} {Energetic} {Fermi} {Millisecond} {Pulsars}},
	volume = {769},
	issn = {0004-637X},
	url = {https://ui.adsabs.harvard.edu/abs/2013ApJ...769..108B},
	doi = {10.1088/0004-637X/769/2/108},
	urldate = {2025-05-06},
	journal = {ApJ},
	author = {Breton, R. P. and van Kerkwijk, M. H. and Roberts, M. S. E. and Hessels, J. W. T. and Camilo, F. and McLaughlin, M. A. and Ransom, S. M. and Ray, P. S. and Stairs, I. H.},
	month = jun,
	year = {2013},
	pages = {108},
}

@article{mata_sanchez_black_2023,
	title = {A black widow population dissection through {HiPERCAM} multiband light-curve modelling},
	volume = {520},
	issn = {0035-8711},
	url = {https://ui.adsabs.harvard.edu/abs/2023MNRAS.520.2217M},
	doi = {10.1093/mnras/stad203},
	urldate = {2025-05-06},
	journal = {MNRAS},
	author = {Mata Sánchez, D. and Kennedy, M. R. and Clark, C. J. and Breton, R. P. and Dhillon, V. S. and Voisin, G. and Camilo, F. and Littlefair, S. and Marsh, T. R. and Stringer, J.},
	month = apr,
	year = {2023},
	pages = {2217--2244},
}

@article{van_der_wateren_irradiated_2022,
	title = {Irradiated but not eclipsed, the case of {PSR} {J0610}−2100},
	volume = {661},
	issn = {0004-6361},
	url = {https://ui.adsabs.harvard.edu/abs/2022A&A...661A..57V},
	doi = {10.1051/0004-6361/202142741},
	urldate = {2025-05-06},
	journal = {A\&A},
	author = {van der Wateren, E. and Bassa, C. G. and Clark, C. J. and Breton, R. P. and Cognard, I. and Guillemot, L. and Janssen, G. H. and Lyne, A. G. and Stappers, B. W. and Theureau, G.},
	month = may,
	year = {2022},
	pages = {A57},
}

@article{kaplan_dense_2018,
	title = {A {Dense} {Companion} to the {Short}-period {Millisecond} {Pulsar} {Binary} {PSR} {J0636}+5128},
	volume = {864},
	issn = {0004-637X},
	url = {https://ui.adsabs.harvard.edu/abs/2018ApJ...864...15K},
	doi = {10.3847/1538-4357/aad54c},
	urldate = {2025-05-06},
	journal = {ApJ},
	author = {Kaplan, D. L. and Stovall, K. and van Kerkwijk, M. H. and Fremling, C. and Istrate, A. G.},
	month = sep,
	year = {2018},
	pages = {15},
}

@article{clark_neutron_2023,
	title = {Neutron star mass estimates from gamma-ray eclipses in spider millisecond pulsar binaries},
	volume = {7},
	issn = {2397-3366},
	url = {https://ui.adsabs.harvard.edu/abs/2023NatAs...7..451C},
	doi = {10.1038/s41550-022-01874-x},
	urldate = {2025-05-06},
	journal = {Nat. Astron.},
	author = {Clark, C. J. and Kerr, M. and Barr, E. D. and Bhattacharyya, B. and Breton, R. P. and Bruel, P. and Camilo, F. and Chen, W. and Cognard, I. and Cromartie, H. T. and Deneva, J. and Dhillon, V. S. and Guillemot, L. and Kennedy, M. R. and Kramer, M. and Lyne, A. G. and Mata Sánchez, D. and Nieder, L. and Phillips, C. and Ransom, S. M. and Ray, P. S. and Roberts, M. S. E. and Roy, J. and Smith, D. A. and Spiewak, R. and Stappers, B. W. and Tabassum, S. and Theureau, G. and Voisin, G.},
	month = apr,
	year = {2023},
	pages = {451--462},
}

@article{li_optical_2014,
	title = {Optical {Counterparts} of {Two} {Fermi} {Millisecond} {Pulsars}: {PSR} {J1301}+0833 and {PSR} {J1628}-3205},
	volume = {795},
	issn = {0004-637X},
	shorttitle = {Optical {Counterparts} of {Two} {Fermi} {Millisecond} {Pulsars}},
	url = {https://ui.adsabs.harvard.edu/abs/2014ApJ...795..115L},
	doi = {10.1088/0004-637X/795/2/115},
	urldate = {2025-05-06},
	journal = {ApJ},
	author = {Li, Miao and Halpern, Jules P. and Thorstensen, John R.},
	month = nov,
	year = {2014},
	pages = {115},
}

@article{reardon_timing_2016,
	title = {Timing analysis for 20 millisecond pulsars in the {Parkes} {Pulsar} {Timing} {Array}},
	volume = {455},
	issn = {0035-8711},
	url = {https://ui.adsabs.harvard.edu/abs/2016MNRAS.455.1751R},
	doi = {10.1093/mnras/stv2395},
	urldate = {2025-05-06},
	journal = {MNRAS},
	author = {Reardon, D. J. and Hobbs, G. and Coles, W. and Levin, Y. and Keith, M. J. and Bailes, M. and Bhat, N. D. R. and Burke-Spolaor, S. and Dai, S. and Kerr, M. and Lasky, P. D. and Manchester, R. N. and Osłowski, S. and Ravi, V. and Shannon, R. M. and van Straten, W. and Toomey, L. and Wang, J. and Wen, L. and You, X. P. and Zhu, X. -J.},
	month = jan,
	year = {2016},
	pages = {1751--1769},
}

@article{vigeland_bayesian_2014,
	title = {Bayesian inference for pulsar-timing models},
	volume = {440},
	issn = {0035-8711},
	url = {https://ui.adsabs.harvard.edu/abs/2014MNRAS.440.1446V},
	doi = {10.1093/mnras/stu312},
	urldate = {2025-05-06},
	journal = {MNRAS},
	author = {Vigeland, Sarah J. and Vallisneri, Michele},
	month = may,
	year = {2014},
	pages = {1446--1457},
}

@article{fonseca_nanograv_2016,
	title = {The {NANOGrav} {Nine}-year {Data} {Set}: {Mass} and {Geometric} {Measurements} of {Binary} {Millisecond} {Pulsars}},
	volume = {832},
	issn = {0004-637X},
	shorttitle = {The {NANOGrav} {Nine}-year {Data} {Set}},
	url = {https://ui.adsabs.harvard.edu/abs/2016ApJ...832..167F},
	doi = {10.3847/0004-637X/832/2/167},
	urldate = {2025-05-06},
	journal = {ApJ},
	author = {Fonseca, Emmanuel and Pennucci, Timothy T. and Ellis, Justin A. and Stairs, Ingrid H. and Nice, David J. and Ransom, Scott M. and Demorest, Paul B. and Arzoumanian, Zaven and Crowter, Kathryn and Dolch, Timothy and Ferdman, Robert D. and Gonzalez, Marjorie E. and Jones, Glenn and Jones, Megan L. and Lam, Michael T. and Levin, Lina and McLaughlin, Maura A. and Stovall, Kevin and Swiggum, Joseph K. and Zhu, Weiwei},
	month = dec,
	year = {2016},
	pages = {167},
}

@inproceedings{gendreau_neutron_2012,
	title = {The {Neutron} star {Interior} {Composition} {ExploreR} ({NICER}): an {Explorer} mission of opportunity for soft x-ray timing spectroscopy},
	volume = {8443},
	shorttitle = {The {Neutron} star {Interior} {Composition} {ExploreR} ({NICER})},
	url = {https://ui.adsabs.harvard.edu/abs/2012SPIE.8443E..13G},
	doi = {10.1117/12.926396},
	urldate = {2025-05-30},
	booktitle = {Space {Telescopes} and {Instrumentation} 2012: {Ultraviolet} to {Gamma} {Ray}},
	author = {Gendreau, Keith C. and Arzoumanian, Zaven and Okajima, Takashi},
	month = sep,
	year = {2012},
	pages = {844313},
}

@article{laycock_pulse-profile_2025,
	title = {Pulse-profile {Modeling} and {Spin}–{Orbit} {Alignment} in a {Suzaku} {Sample} of {Accreting} {X}-{Ray} {Binary} {Pulsars}},
	volume = {978},
	issn = {0004-637X},
	url = {https://ui.adsabs.harvard.edu/abs/2025ApJ...978...80L},
	doi = {10.3847/1538-4357/ad9399},
	urldate = {2025-06-11},
	journal = {ApJ},
	author = {Laycock, Silas G. T. and Cappallo, Rigel C. and Pradhan, Pragati and Christodoulou, Dimitris M. and Paul, Biswajit},
	month = jan,
	year = {2025},
	pages = {80},
}

@article{abolmasov_spin_2024,
	title = {Spin {Evolution} of {Neutron} {Stars}},
	volume = {12},
	url = {https://ui.adsabs.harvard.edu/abs/2024Galax..12....7A},
	doi = {10.3390/galaxies12010007},
	urldate = {2025-07-16},
	journal = {Galaxies},
	author = {Abolmasov, Pavel and Biryukov, Anton and Popov, Sergei B.},
	month = feb,
	year = {2024},
	pages = {7},
}

@article{choudhury_nicer_2024,
	title = {A {NICER} {View} of the {Nearest} and {Brightest} {Millisecond} {Pulsar}: {PSR} {J0437}–4715},
	volume = {971},
	issn = {0004-637X},
	shorttitle = {A {NICER} {View} of the {Nearest} and {Brightest} {Millisecond} {Pulsar}},
	url = {https://ui.adsabs.harvard.edu/abs/2024ApJ...971L..20C},
	doi = {10.3847/2041-8213/ad5a6f},
	urldate = {2025-07-16},
	journal = {ApJ},
	author = {Choudhury, Devarshi and Salmi, Tuomo and Vinciguerra, Serena and Riley, Thomas E. and Kini, Yves and Watts, Anna L. and Dorsman, Bas and Bogdanov, Slavko and Guillot, Sebastien and Ray, Paul S. and Reardon, Daniel J. and Remillard, Ronald A. and Bilous, Anna V. and Huppenkothen, Daniela and Lattimer, James M. and Rutherford, Nathan and Arzoumanian, Zaven and Gendreau, Keith C. and Morsink, Sharon M. and Ho, Wynn C. G.},
	month = aug,
	year = {2024},
	pages = {L20},
}

@article{salmi_radius_2024,
	title = {The {Radius} of the {High}-mass {Pulsar} {PSR} {J0740}+6620 with 3.6 yr of {NICER} {Data}},
	volume = {974},
	issn = {0004-637X},
	url = {https://ui.adsabs.harvard.edu/abs/2024ApJ...974..294S},
	doi = {10.3847/1538-4357/ad5f1f},
	urldate = {2025-07-16},
	journal = {ApJ},
	author = {Salmi, Tuomo and Choudhury, Devarshi and Kini, Yves and Riley, Thomas E. and Vinciguerra, Serena and Watts, Anna L. and Wolff, Michael T. and Arzoumanian, Zaven and Bogdanov, Slavko and Chakrabarty, Deepto and Gendreau, Keith and Guillot, Sebastien and Ho, Wynn C. G. and Huppenkothen, Daniela and Ludlam, Renee M. and Morsink, Sharon M. and Ray, Paul S.},
	month = oct,
	year = {2024},
	pages = {294},
}

@article{yang_magnetic_2023,
	title = {Magnetic {Inclination} {Evolution} of {Accreting} {Neutron} {Stars} in {Intermediate}/{Low}-mass {X}-{Ray} {Binaries}},
	volume = {945},
	issn = {0004-637X},
	url = {https://ui.adsabs.harvard.edu/abs/2023ApJ...945....2Y},
	doi = {10.3847/1538-4357/acba09},
	urldate = {2025-07-16},
	journal = {ApJ},
	author = {Yang, Hao-ran and Li, Xiang-dong},
	month = mar,
	year = {2023},
	pages = {2},
}

@article{kirk_pulsed_2002,
	title = {Pulsed radiation from neutron star winds},
	volume = {388},
	issn = {0004-6361},
	url = {https://ui.adsabs.harvard.edu/abs/2002A&A...388L..29K},
	doi = {10.1051/0004-6361:20020599},
	urldate = {2025-07-18},
	journal = {A\&A},
	author = {Kirk, J. G. and Skjæraasen, O. and Gallant, Y. A.},
	month = jun,
	year = {2002},
	pages = {L29--L32},
}

@article{polzin_study_2020,
	title = {Study of spider pulsar binary eclipses and discovery of an eclipse mechanism transition},
	volume = {494},
	issn = {0035-8711},
	url = {https://ui.adsabs.harvard.edu/abs/2020MNRAS.494.2948P},
	doi = {10.1093/mnras/staa596},
	urldate = {2025-07-18},
	journal = {MNRAS},
	author = {Polzin, E. J. and Breton, R. P. and Bhattacharyya, B. and Scholte, D. and Sobey, C. and Stappers, B. W.},
	month = may,
	year = {2020},
	pages = {2948--2968},
}

@article{petri_unified_2011,
	title = {A unified polar cap/striped wind model for pulsed radio and gamma-ray emission in pulsars},
	volume = {412},
	issn = {0035-8711},
	url = {https://ui.adsabs.harvard.edu/abs/2011MNRAS.412.1870P},
	doi = {10.1111/j.1365-2966.2010.18023.x},
	urldate = {2025-07-18},
	journal = {MNRAS},
	author = {Pétri, J.},
	month = apr,
	year = {2011},
	pages = {1870--1880},
}

@article{petri_pulsar_2012,
	title = {The pulsar force-free magnetosphere linked to its striped wind: time-dependent pseudo-spectral simulations},
	volume = {424},
	issn = {0035-8711},
	shorttitle = {The pulsar force-free magnetosphere linked to its striped wind},
	url = {https://ui.adsabs.harvard.edu/abs/2012MNRAS.424..605P},
	doi = {10.1111/j.1365-2966.2012.21238.x},
	urldate = {2025-07-18},
	journal = {MNRAS},
	author = {Pétri, J.},
	month = jul,
	year = {2012},
	pages = {605--619},
}

@article{petri_young_2021,
	title = {Young radio-loud gamma-ray pulsar light curve fitting},
	volume = {654},
	issn = {0004-6361},
	url = {https://ui.adsabs.harvard.edu/abs/2021A&A...654A.106P},
	doi = {10.1051/0004-6361/202141272},
	urldate = {2025-07-18},
	journal = {A\&A},
	author = {Pétri, J. and Mitra, D.},
	month = oct,
	year = {2021},
	pages = {A106},
}

@article{riley_nicer_2019,
	title = {A {NICER} {View} of {PSR} {J0030}+0451: {Millisecond} {Pulsar} {Parameter} {Estimation}},
	volume = {887},
	issn = {0004-637X},
	shorttitle = {A {NICER} {View} of {PSR} {J0030}+0451},
	url = {https://ui.adsabs.harvard.edu/abs/2019ApJ...887L..21R},
	doi = {10.3847/2041-8213/ab481c},
	urldate = {2025-07-18},
	journal = {ApJ},
	author = {Riley, T. E. and Watts, A. L. and Bogdanov, S. and Ray, P. S. and Ludlam, R. M. and Guillot, S. and Arzoumanian, Z. and Baker, C. L. and Bilous, A. V. and Chakrabarty, D. and Gendreau, K. C. and Harding, A. K. and Ho, W. C. G. and Lattimer, J. M. and Morsink, S. M. and Strohmayer, T. E.},
	month = dec,
	year = {2019},
	pages = {L21},
}

@book{press_numerical_2007,
	title = {Numerical {Recipes} 3rd {Edition} : {The} {Art} of {Scientific} {Computing}},
	author = {Press, W. H. and Teukolsky, S. A. and Vetterling, W. T. and Flannery, B. P.},
	year = {2007},
}

@article{draghis_multiband_2019,
	title = {Multiband {Optical} {Light} {Curves} of {Black}-widow {Pulsars}},
	volume = {883},
	doi = {10.3847/1538-4357/ab378b},
	number = {1},
	journal = {ApJ},
	author = {Draghis, Paul and Romani, Roger W. and Filippenko, Alexei V. and Brink, Thomas G. and Zheng, WeiKang and Halpern, Jules P. and Camilo, Fernando},
	month = sep,
	year = {2019},
	pages = {108},
}

@article{petri_multi-wavelength_2024,
	title = {Multi-wavelength pulse profiles from the force-free neutron star magnetosphere},
	volume = {687},
	issn = {0004-6361},
	url = {https://ui.adsabs.harvard.edu/abs/2024A&A...687A.169P},
	doi = {10.1051/0004-6361/202348069},
	urldate = {2025-07-18},
	journal = {A\&A},
	author = {Pétri, J.},
	month = jul,
	year = {2024},
	pages = {A169},
}

@article{ashton_bilby_2019,
	title = {{BILBY}: {A} {User}-friendly {Bayesian} {Inference} {Library} for {Gravitational}-wave {Astronomy}},
	volume = {241},
	issn = {0067-0049},
	shorttitle = {{BILBY}},
	url = {https://ui.adsabs.harvard.edu/abs/2019ApJS..241...27A},
	doi = {10.3847/1538-4365/ab06fc},
	urldate = {2025-07-18},
	journal = {ApJS},
	author = {Ashton, Gregory and Hübner, Moritz and Lasky, Paul D. and Talbot, Colm and Ackley, Kendall and Biscoveanu, Sylvia and Chu, Qi and Divakarla, Atul and Easter, Paul J. and Goncharov, Boris and Hernandez Vivanco, Francisco and Harms, Jan and Lower, Marcus E. and Meadors, Grant D. and Melchor, Denyz and Payne, Ethan and Pitkin, Matthew D. and Powell, Jade and Sarin, Nikhil and Smith, Rory J. E. and Thrane, Eric},
	month = apr,
	year = {2019},
	pages = {27},
}

@article{tauris_formation_2017,
	title = {Formation of {Double} {Neutron} {Star} {Systems}},
	volume = {846},
	issn = {0004-637X},
	url = {https://ui.adsabs.harvard.edu/abs/2017ApJ...846..170T},
	doi = {10.3847/1538-4357/aa7e89},
	urldate = {2025-07-18},
	journal = {ApJ},
	author = {Tauris, T. M. and Kramer, M. and Freire, P. C. C. and Wex, N. and Janka, H. -T. and Langer, N. and Podsiadlowski, Ph. and Bozzo, E. and Chaty, S. and Kruckow, M. U. and van den Heuvel, E. P. J. and Antoniadis, J. and Breton, R. P. and Champion, D. J.},
	month = sep,
	year = {2017},
	pages = {170},
}

@article{spera_very_2017,
	title = {Very massive stars, pair-instability supernovae and intermediate-mass black holes with the sevn code},
	volume = {470},
	issn = {0035-8711},
	url = {https://ui.adsabs.harvard.edu/abs/2017MNRAS.470.4739S},
	doi = {10.1093/mnras/stx1576},
	urldate = {2025-07-18},
	journal = {MNRAS},
	author = {Spera, Mario and Mapelli, Michela},
	month = oct,
	year = {2017},
	pages = {4739--4749},
}

@article{iorio_compact_2023,
	title = {Compact object mergers: exploring uncertainties from stellar and binary evolution with {SEVN}},
	volume = {524},
	issn = {0035-8711},
	shorttitle = {Compact object mergers},
	url = {https://ui.adsabs.harvard.edu/abs/2023MNRAS.524..426I},
	doi = {10.1093/mnras/stad1630},
	urldate = {2025-07-18},
	journal = {MNRAS},
	author = {Iorio, Giuliano and Mapelli, Michela and Costa, Guglielmo and Spera, Mario and Escobar, Gastón J. and Sgalletta, Cecilia and Trani, Alessandro A. and Korb, Erika and Santoliquido, Filippo and Dall'Amico, Marco and Gaspari, Nicola and Bressan, Alessandro},
	month = sep,
	year = {2023},
	pages = {426--470},
}

@article{sautron_galactic_2024,
	title = {The {Galactic} population of canonical pulsars: {II}. {Taking} into account several evolutionary paths: {Interstellar} medium interaction, {Galactic} potential, and a death valley},
	volume = {691},
	issn = {0004-6361},
	shorttitle = {The {Galactic} population of canonical pulsars},
	url = {https://ui.adsabs.harvard.edu/abs/2024A&A...691A.349S},
	doi = {10.1051/0004-6361/202451097},
	urldate = {2025-07-18},
	journal = {A\&A},
	author = {Sautron, Mattéo and Pétri, Jérôme and Mitra, Dipanjan and Dirson, Ludmilla},
	month = nov,
	year = {2024},
	pages = {A349},
}

@article{fonseca_refined_2021,
	title = {Refined {Mass} and {Geometric} {Measurements} of the {High}-mass {PSR} {J0740}+6620},
	volume = {915},
	issn = {0004-637X},
	url = {https://ui.adsabs.harvard.edu/abs/2021ApJ...915L..12F},
	doi = {10.3847/2041-8213/ac03b8},
	urldate = {2025-07-23},
	journal = {ApJ},
	author = {Fonseca, E. and Cromartie, H. T. and Pennucci, T. T. and Ray, P. S. and Kirichenko, A. Yu. and Ransom, S. M. and Demorest, P. B. and Stairs, I. H. and Arzoumanian, Z. and Guillemot, L. and Parthasarathy, A. and Kerr, M. and Cognard, I. and Baker, P. T. and Blumer, H. and Brook, P. R. and DeCesar, M. and Dolch, T. and Dong, F. A. and Ferrara, E. C. and Fiore, W. and Garver-Daniels, N. and Good, D. C. and Jennings, R. and Jones, M. L. and Kaspi, V. M. and Lam, M. T. and Lorimer, D. R. and Luo, J. and McEwen, A. and McKee, J. W. and McLaughlin, M. A. and McMann, N. and Meyers, B. W. and Naidu, A. and Ng, C. and Nice, D. J. and Pol, N. and Radovan, H. A. and Shapiro-Albert, B. and Tan, C. M. and Tendulkar, S. P. and Swiggum, J. K. and Wahl, H. M. and Zhu, W. W.},
	month = jul,
	year = {2021},
	pages = {L12},
}

@article{miller_psr_2019,
	title = {{PSR} {J0030}+0451 {Mass} and {Radius} from {NICER} {Data} and {Implications} for the {Properties} of {Neutron} {Star} {Matter}},
	volume = {887},
	issn = {0004-637X},
	url = {https://ui.adsabs.harvard.edu/abs/2019ApJ...887L..24M},
	doi = {10.3847/2041-8213/ab50c5},
	urldate = {2025-10-24},
	journal = {ApJ},
	author = {Miller, M. C. and Lamb, F. K. and Dittmann, A. J. and Bogdanov, S. and Arzoumanian, Z. and Gendreau, K. C. and Guillot, S. and Harding, A. K. and Ho, W. C. G. and Lattimer, J. M. and Ludlam, R. M. and Mahmoodifar, S. and Morsink, S. M. and Ray, P. S. and Strohmayer, T. E. and Wood, K. S. and Enoto, T. and Foster, R. and Okajima, T. and Prigozhin, G. and Soong, Y.},
	month = dec,
	year = {2019},
	pages = {L24},
}

@article{miller_radius_2021,
	title = {The {Radius} of {PSR} {J0740}+6620 from {NICER} and {XMM}-{Newton} {Data}},
	volume = {918},
	issn = {0004-637X},
	url = {https://ui.adsabs.harvard.edu/abs/2021ApJ...918L..28M},
	doi = {10.3847/2041-8213/ac089b},
	urldate = {2025-10-24},
	journal = {ApJ},
	author = {Miller, M. C. and Lamb, F. K. and Dittmann, A. J. and Bogdanov, S. and Arzoumanian, Z. and Gendreau, K. C. and Guillot, S. and Ho, W. C. G. and Lattimer, J. M. and Loewenstein, M. and Morsink, S. M. and Ray, P. S. and Wolff, M. T. and Baker, C. L. and Cazeau, T. and Manthripragada, S. and Markwardt, C. B. and Okajima, T. and Pollard, S. and Cognard, I. and Cromartie, H. T. and Fonseca, E. and Guillemot, L. and Kerr, M. and Parthasarathy, A. and Pennucci, T. T. and Ransom, S. and Stairs, I.},
	month = sep,
	year = {2021},
	pages = {L28},
}

@article{dittmann_more_2024,
	title = {A {More} {Precise} {Measurement} of the {Radius} of {PSR} {J0740}+6620 {Using} {Updated} {NICER} {Data}},
	volume = {974},
	issn = {0004-637X},
	url = {https://ui.adsabs.harvard.edu/abs/2024ApJ...974..295D},
	doi = {10.3847/1538-4357/ad5f1e},
	urldate = {2025-10-24},
	journal = {ApJ},
	author = {Dittmann, Alexander J. and Miller, M. Coleman and Lamb, Frederick K. and Holt, Isiah M. and Chirenti, Cecilia and Wolff, Michael T. and Bogdanov, Slavko and Guillot, Sebastien and Ho, Wynn C. G. and Morsink, Sharon M. and Arzoumanian, Zaven and Gendreau, Keith C.},
	month = oct,
	year = {2024},
	pages = {295},
}

@article{contopoulos_pulsar_2010,
	title = {The pulsar synchrotron in {3D}: curvature radiation},
	volume = {404},
	issn = {0035-8711},
	shorttitle = {The pulsar synchrotron in {3D}},
	url = {https://ui.adsabs.harvard.edu/abs/2010MNRAS.404..767C},
	doi = {10.1111/j.1365-2966.2010.16338.x},
	urldate = {2025-10-24},
	journal = {MNRAS},
	author = {Contopoulos, Ioannis and Kalapotharakos, Constantinos},
	month = may,
	year = {2010},
	pages = {767--778},
}

@article{bai_modeling_2010,
	title = {Modeling of {Gamma}-ray {Pulsar} {Light} {Curves} {Using} the {Force}-free {Magnetic} {Field}},
	volume = {715},
	issn = {0004-637X},
	url = {https://ui.adsabs.harvard.edu/abs/2010ApJ...715.1282B},
	doi = {10.1088/0004-637X/715/2/1282},
	urldate = {2025-10-24},
	journal = {ApJ},
	author = {Bai, Xue-Ning and Spitkovsky, Anatoly},
	month = jun,
	year = {2010},
	pages = {1282--1301},
}

@article{kalapotharakos_gamma-ray_2014,
	title = {Gamma-{Ray} {Emission} in {Dissipative} {Pulsar} {Magnetospheres}: {From} {Theory} to {Fermi} {Observations}},
	volume = {793},
	issn = {0004-637X},
	shorttitle = {Gamma-{Ray} {Emission} in {Dissipative} {Pulsar} {Magnetospheres}},
	url = {https://ui.adsabs.harvard.edu/abs/2014ApJ...793...97K},
	doi = {10.1088/0004-637X/793/2/97},
	urldate = {2025-10-24},
	journal = {ApJ},
	author = {Kalapotharakos, Constantinos and Harding, Alice K. and Kazanas, Demosthenes},
	month = oct,
	year = {2014},
	pages = {97},
}

@article{cerutti_synthetic_2025,
	title = {Synthetic pulsar light curves from global kinetic simulations and comparison with the {Fermi}-{LAT} catalog},
	volume = {695},
	issn = {0004-6361},
	url = {https://ui.adsabs.harvard.edu/abs/2025A&A...695A..93C},
	doi = {10.1051/0004-6361/202451948},
	urldate = {2025-10-24},
	journal = {A\&A},
	author = {Cerutti, Benoît and Figueiredo, Enzo and Dubus, Guillaume},
	month = mar,
	year = {2025},
	pages = {A93},
}

@article{cerutti_modelling_2016,
	title = {Modelling high-energy pulsar light curves from first principles},
	volume = {457},
	issn = {0035-8711},
	url = {https://ui.adsabs.harvard.edu/abs/2016MNRAS.457.2401C},
	doi = {10.1093/mnras/stw124},
	urldate = {2025-10-24},
	journal = {MNRAS},
	author = {Cerutti, Benoît and Philippov, Alexander A. and Spitkovsky, Anatoly},
	month = apr,
	year = {2016},
	pages = {2401--2414},
}

@article{kalapotharakos_three-dimensional_2018,
	title = {Three-dimensional {Kinetic} {Pulsar} {Magnetosphere} {Models}: {Connecting} to {Gamma}-{Ray} {Observations}},
	volume = {857},
	issn = {0004-637X},
	shorttitle = {Three-dimensional {Kinetic} {Pulsar} {Magnetosphere} {Models}},
	url = {https://ui.adsabs.harvard.edu/abs/2018ApJ...857...44K},
	doi = {10.3847/1538-4357/aab550},
	urldate = {2025-10-24},
	journal = {ApJ},
	author = {Kalapotharakos, Constantinos and Brambilla, Gabriele and Timokhin, Andrey and Harding, Alice K. and Kazanas, Demosthenes},
	month = apr,
	year = {2018},
	pages = {44},
}

@article{kalapotharakos_gamma-ray_2023,
	title = {The {Gamma}-{Ray} {Pulsar} {Phenomenology} in {View} of {3D} {Kinetic} {Global} {Magnetosphere} {Models}},
	volume = {954},
	issn = {0004-637X},
	url = {https://ui.adsabs.harvard.edu/abs/2023ApJ...954..204K},
	doi = {10.3847/1538-4357/ace972},
	urldate = {2025-10-24},
	journal = {ApJ},
	author = {Kalapotharakos, Constantinos and Wadiasingh, Zorawar and Harding, Alice K. and Kazanas, Demosthenes},
	month = sep,
	year = {2023},
	pages = {204},
}

@article{philippov_ab-initio_2018,
	title = {Ab-initio {Pulsar} {Magnetosphere}: {Particle} {Acceleration} in {Oblique} {Rotators} and {High}-energy {Emission} {Modeling}},
	volume = {855},
	issn = {0004-637X},
	shorttitle = {Ab-initio {Pulsar} {Magnetosphere}},
	url = {https://ui.adsabs.harvard.edu/abs/2018ApJ...855...94P},
	doi = {10.3847/1538-4357/aaabbc},
	urldate = {2025-10-24},
	journal = {ApJ},
	author = {Philippov, Alexander A. and Spitkovsky, Anatoly},
	month = mar,
	year = {2018},
	pages = {94},
}

@article{cao_modeling_2019,
	title = {Modeling {Gamma}-{Ray} {Light} {Curves} with {More} {Realistic} {Pulsar} {Magnetospheres}},
	volume = {874},
	issn = {0004-637X},
	url = {https://ui.adsabs.harvard.edu/abs/2019ApJ...874..166C},
	doi = {10.3847/1538-4357/ab0d20},
	urldate = {2025-10-24},
	journal = {ApJ},
	author = {Cao, Gang and Yang, Xiongbang},
	month = apr,
	year = {2019},
	pages = {166},
}

@article{cao_pulsar_2022,
	title = {The {Pulsar} {Gamma}-{Ray} {Emission} from {High}-resolution {Dissipative} {Magnetospheres}},
	volume = {925},
	issn = {0004-637X},
	url = {https://ui.adsabs.harvard.edu/abs/2022ApJ...925..130C},
	doi = {10.3847/1538-4357/ac3dea},
	urldate = {2025-10-24},
	journal = {ApJ},
	author = {Cao, Gang and Yang, Xiongbang},
	month = feb,
	year = {2022},
	pages = {130},
}

@article{lohmer_shapiro_2005,
	title = {Shapiro {Delay} in the {PSR} {J1640}+2224 {Binary} {System}},
	volume = {621},
	issn = {0004-637X},
	url = {https://ui.adsabs.harvard.edu/abs/2005ApJ...621..388L},
	doi = {10.1086/427404},
	urldate = {2025-11-17},
	journal = {ApJ},
	author = {Löhmer, Oliver and Lewandowski, Wojciech and Wolszczan, Alex and Wielebinski, Richard},
	month = mar,
	year = {2005},
	pages = {388--392},
}

@article{blanchard_census_2025,
	title = {A census of galactic spider binary millisecond pulsars with the {Nançay} {Radio} {Telescope}},
	volume = {698},
	copyright = {https://creativecommons.org/licenses/by/4.0},
	issn = {0004-6361, 1432-0746},
	url = {https://www.aanda.org/10.1051/0004-6361/202453499},
	doi = {10.1051/0004-6361/202453499},
	language = {en},
	urldate = {2025-12-04},
	journal = {A\&A},
	author = {Blanchard, C. and Guillemot, L. and Voisin, G. and Cognard, I. and Theureau, G.},
	month = jun,
	year = {2025},
	pages = {A239},
}

@article{lorimer_binary_2008,
	title = {Binary and {Millisecond} {Pulsars}},
	volume = {11},
	issn = {1433-8351},
	url = {https://doi.org/10.12942/lrr-2008-8},
	doi = {10.12942/lrr-2008-8},
	language = {en},
	number = {1},
	urldate = {2025-12-04},
	journal = {Living Rev. Relativ.},
	author = {Lorimer, Duncan R.},
	month = nov,
	year = {2008},
	pages = {8},
}

@misc{sautron_born_2025,
	
	shorttitle = {Born to be recycled},
	url = {http://arxiv.org/abs/2510.15661},
	doi = {10.48550/arXiv.2510.15661},
	urldate = {2025-12-04},
	publisher = {arXiv},
	author = {Sautron, Mattéo and Pétri, Jérôme and Mitra, Dipanjan and Dupuy--Junet, Adélie and Pietrin, Marie-Eloïse},
	month = oct,
	year = {2025},
	note = {arXiv:2510.15661},
}

@misc{andrae_dos_2010,
	
	url = {http://arxiv.org/abs/1012.3754},
	doi = {10.48550/arXiv.1012.3754},
	urldate = {2025-12-04},
	publisher = {arXiv},
	author = {Andrae, Rene and Schulze-Hartung, Tim and Melchior, Peter},
	month = dec,
	year = {2010},
	note = {arXiv:1012.3754},
}

@article{hui_high_2019,
	title = {High {Energy} {Radiation} from {Spider} {Pulsars}},
	volume = {7},
	copyright = {http://creativecommons.org/licenses/by/3.0/},
	issn = {2075-4434},
	url = {https://www.mdpi.com/2075-4434/7/4/93},
	doi = {10.3390/galaxies7040093},
	language = {en},
	number = {4},
	urldate = {2025-12-04},
	journal = {Galaxies},
	author = {Hui, Chung Yue and Li, Kwan Lok},
	month = dec,
	year = {2019},
	pages = {93},
}

@misc{kramer_radio_2025,
	title = {Radio emission from beyond the light cylinder in millisecond pulsars},
	url = {http://arxiv.org/abs/2510.05778},
	doi = {10.48550/arXiv.2510.05778},
	urldate = {2025-12-04},
	publisher = {arXiv},
	author = {Kramer, Michael and Johnston, Simon},
	month = oct,
	year = {2025},
	note = {arXiv:2510.05778},
}

@article{kramer_strong-field_2021,
	title = {Strong-{Field} {Gravity} {Tests} with the {Double} {Pulsar}},
	volume = {11},
	url = {https://link.aps.org/doi/10.1103/PhysRevX.11.041050},
	doi = {10.1103/PhysRevX.11.041050},
	number = {4},
	urldate = {2025-12-04},
	journal = {Phys. Rev. X},
	author = {Kramer, M. and Stairs, I. H. and Manchester, R. N. and Wex, N. and Deller, A. T. and Coles, W. A. and Ali, M. and Burgay, M. and Camilo, F. and Cognard, I. and Damour, T. and Desvignes, G. and Ferdman, R. D. and Freire, P. C. C. and Grondin, S. and Guillemot, L. and Hobbs, G. B. and Janssen, G. and Karuppusamy, R. and Lorimer, D. R. and Lyne, A. G. and McKee, J. W. and McLaughlin, M. and Münch, L. E. and Perera, B. B. P. and Pol, N. and Possenti, A. and Sarkissian, J. and Stappers, B. W. and Theureau, G.},
	month = dec,
	year = {2021},
	note = {Publisher: American Physical Society},
	pages = {041050},
}

@article{lewis1995fast,
  author    = {Lewis, J. P.},
  title     = {Fast Normalized Cross-Correlation},
  journal   = {Vision Interface},
  volume    = {},
  number    = {},
  pages     = {120--123},
  year      = {1995},
  note      = {Industrial Light \& Magic},
  url       = {http://scribblethink.org/Work/nvisionInterface/nip.pdf}
}

@article{jacovitti1993normalized,
  author    = {Jacovitti, G. and Scarano, G.},
  title     = {A normalized correlation-based algorithm for time delay estimation},
  journal   = {IEEE Transactions on Signal Processing},
  volume    = {41},
  number    = {1},
  pages     = {417--421},
  year      = {1993},
  doi       = {10.1109/78.193228},
  issn      = {1053-587X}
}

\appendix

\section{\cor{Fitting process}}\label{appendix:fittingprocess}

In this section, we explain how to analyse an observed $\gamma$-ray light curve obtained from the Fermi/LAT telescope. Such curve is a discrete signal $u=(u[k], k=0, \dots, N-1)$, each measurement $u[k]$ being associated with a corresponding phase $\varphi[k]={k}/{N}$. The Fermi data set provides $N$ measurement points, with the number of points $N$ varying depending on the luminosity of each pulsar. For the brightest pulsars of our sample, $N$ can reach $100$ but can also go below $30$ for faint pulsars with poor photon statistics.

\subsection{Fitting method}
In previous sections, we explained that the shape of such a signal depends on the geometry of the pulsar which is described by a couple of angles $\theta=(\chi,\zeta)$; to be admissible the parameter $\theta$ must belongs to:
\begin{equation}
    \Theta =\{(\chi,\zeta) \in [0,\frac{\pi}{2}]^2 :  |\zeta - \frac{\pi}{2}| \le \chi \text{ and } |\zeta - \chi| \le \rho   \}
\end{equation}
with $\rho$ the half-opening angle of the radio beam cone. 




For each parameter $\theta=(\chi,\zeta)$, a modelling continuous signal $V^\theta(\varphi)$ is available for analysis. We denote by $v^\theta$ its discretisation $v^\theta[k]= V^\theta(\varphi_k)$.



Remark: Our numerical process  gives the  function $\varphi \to V^\theta(\varphi )$  at some phases that are not the $\varphi_k$. An interpolation via Fourier series allows us to evaluate it precisely at the $\varphi_k$. 


The collection $\{v^\theta:\theta \in \Theta\}$ constitutes the atlas encompassing all possible discrete signal shapes. An illustration of this atlas is given in \cite{petri_multi-wavelength_2024}. The observed signal $u$ must be a noisy variant of one of the signals within this atlas, subject to three  transformations: phase-shift, spatial scaling (intensity normalisation), and spatial translation (background level).

Let's be more precise. For a discrete signal $v$, we denote by $v_t$ its phase-shifted version: $v_t[k]=v[k-t]$ (the minus sign being considered modulo $N$ because of the periodicity of the signal).  Our purpose is to find  parameters: 
\begin{equation}
 \theta=(\chi,\zeta)\in \Theta,\  a>0,\  b\in \mathbb R,\  t\in\mathbb Z   
\end{equation} so that
\begin{equation}
a\, v^{\theta}_t + b \text{ fits closely to } u.    
\end{equation}
We solve this problem in two steps: 

\textbf{First step: } For a given $\theta \in \Theta$, the normalized-cross-correlation methods (see Appendix~\ref{appendix:crosscorrelation}) allows us to find  efficiently the triplet $(\hat t,\hat a,\hat b)$ minimizing $
   || u - (a v^{\theta}_t + b) ||^2
$ .  The fitted signal is denoted by:
\begin{equation}
    f(\theta) =  \hat a \, v^{\theta}_{\hat t} + {\hat b}
\end{equation}
We express $f(k,\theta)=f(\theta)[k]$ for the fitted signal at phase $\varphi[k]$.

\textbf{Second step: }   We suppose that each measure $u[k]$ is obtained by adding to $f(k,\theta)$ a Gaussian noise with  a known standard deviation $\sigma_k$ (known from the Fermi data set).  
So $u[k]$ is viewed as a sample of a random variable with density:
\begin{equation}
     y_k \to p(y_k | \theta) = \frac{1}{\sqrt{2 \, \pi \, \sigma_k^2}} \exp\left( -\frac{(y_k - f(k, \theta))^2}{2 \, \sigma_k^2} \right) \ .
\end{equation}
Assuming that the noises are independent from one measurement to the next, the entire observed signal $u$ is regarded as a sample from a random vector with density:
\begin{equation}
  y \to p(y | \theta) = \prod_{k=0}^{N-1} \frac{1}{\sqrt{2 \, \pi \, \sigma_k^2}} \exp\left( -\frac{(y_k - f(k, \theta))^2}{2 \, \sigma_k^2} \right)
\end{equation}
where $y=(y_0,\dots, y_{N-1})$ represents the generic element of $\mathbb R^N$. Taking the logarithm of the expression above yields
\begin{equation}\label{log-expression}
  \log p({y} | \theta) = -\frac{1}{2} \sum_{k=0}^{N-1}  \log(2 \,\pi \, \sigma_k^2) -\frac{1}{2} \sum_{k=0}^{N-1}\frac{(y_k - f(k, \theta))^2}{\sigma_k^2} 
\end{equation}
Note that the left term  of the difference above does not depend on the parameter $\theta$ and will be ignored during the optimization process below.  

Within the Bayesian framework, it is postulated that the parameter $\theta$ follows a prior distribution denoted as $\pi(\theta)$ (two options for such a distribution are described subsequently). Bayes' theorem provides a formulation for the posterior distribution of $\theta$ given $y$
\begin{equation}
p(\theta | {y}) = \frac{p({y} | \theta) \, \pi(\theta)}{p({y})} \ .
\end{equation}
The denominator term $p(y)= \int_{\Theta} p({y} | \theta) \pi(\theta) \, d\theta$ is referred to as the evidence. Its calculation is irrelevant because it does not depend on $\theta$.

To determine the parameter $\hat{\theta}$ that renders the observed signal $u$ the most realistic among all possible parameters $\theta \in \Theta$, we need to maximize the posterior probability evaluated at $y=u$ so we have to find  $\hat\theta = \text{argmax}_\theta p(\theta | u)$. Considering the log-expression (\ref{log-expression}), the optimization problem can be simplified to
  \begin{equation}
  \hat{\theta} = \arg\min_{\theta\in \Theta}\left[  \frac{1}{2} \sum_{k=0}^{N-1}\frac{(u[k] - f(k, \theta))^2}{\sigma_k^2} - \log \pi(\theta)  \right]  \ .
  \end{equation}
To enhance the importance of measurements at the signal's peak, we introduce a weighting factor to the previously stated expression:
  \begin{equation} \label{final-expression}
  \hat{\theta} = \arg\min_{\theta\in \Theta}\left[  \frac{1}{2} \sum_{k=0}^{N-1}\frac{(u[k]+1)^2 (u[k] - f(k, \theta))^2}{\sigma_k^2} - \log \pi(\theta)   \right]  \ .
  \end{equation}
Observe that the preceding adjustment is equivalent to substituting $\sigma_i$ with $\frac{\sigma_i}{u[i]+1}$; thereby supposing that the measurements on the peaks are less susceptible to noise interference.

Finally, to determine the minimum, a finite grid $\Theta' \subset \Theta$ is selected and evaluate all quantities (\ref{final-expression}) for $\theta \in \Theta'$. This task is facilitated by the fact that $\Theta$ is a simple subset of $\mathbb{R}^2$, and by the computational efficiency of the function $f$,  via the normalized cross-correlation method.




However, for our simple 2D optimization problem, the employed grid method facilitates the rapid identification of the minimum with high precision. We utilize a grid with variations at one-degree increments in $\chi$ and $\zeta$, granularity could be extended if necessary.


\subsection{Application to the Shapiro delay}

We apply the previous process with two different priors:

\textbf{First prior:}  $\pi$ the isotropic  distribution over $\Theta$ (in other word, the uniform distribution of the restricted part of the sphere). 

\textbf{Second prior}: Considering the orbital inclination $i$:
\begin{equation}
    \pi(\theta) = \pi(\chi,\zeta) = {C_\chi}\, e^{-\frac{1}{2 \times 0.01}(\zeta-i)^2} 1_{(\chi,\zeta)\in \Theta}
\end{equation}
with $C_\chi$ a normalization factor so that the density has integral is 1. Note that this constant depends on $\chi$ because the domain is not rectangular. However, for the parameters that we estimate, the Gaussian is peaked enough so that this dependence can be neglected. Thus, this multiplicative constant vanishes during the optimization process.


\subsection{Example of two fitting results}

\begin{figure*}[h]
    \centering
    \subfigure[J0740+6620]{\includegraphics[width=0.45\linewidth]{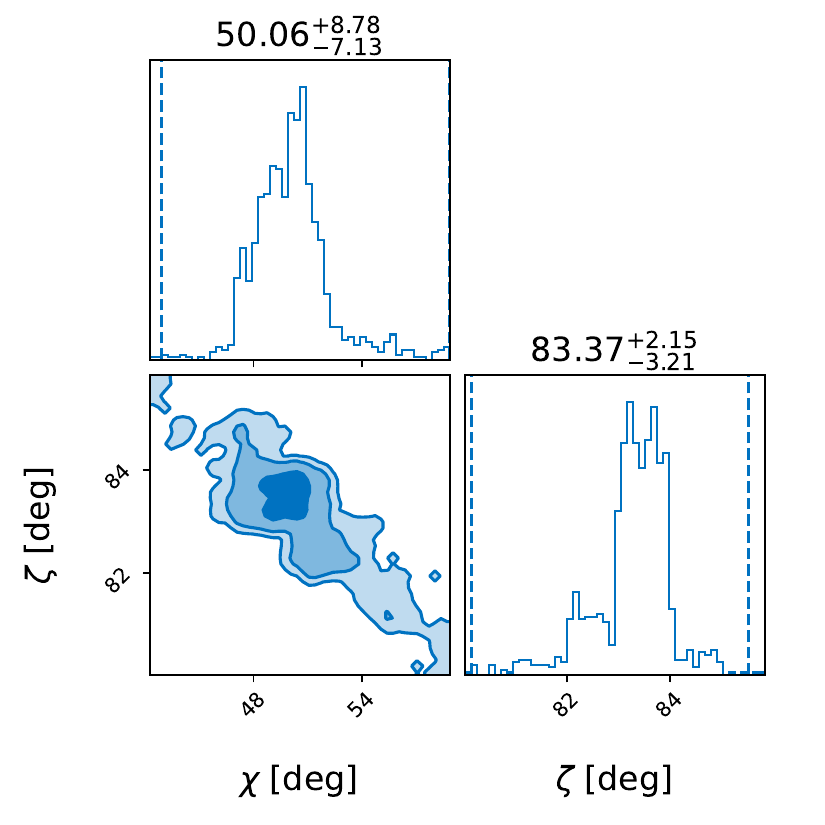}}
    \subfigure[J1514-4946]{\includegraphics[width=0.45\linewidth]{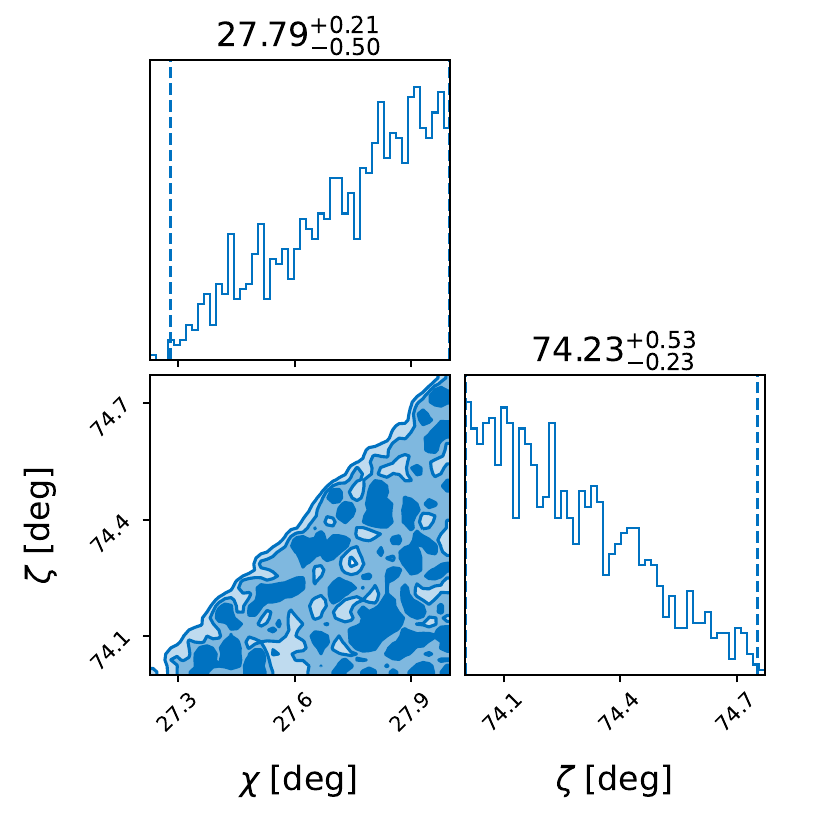}}
    \caption{Corner plots for the fitting in the general case, without Shapiro constraint, for pulsar J0740$+$6620 on the left panel, and for pulsar J1514$-$4946 on the right panel.}
    \label{fig:cornerplot}
\end{figure*}

The fitting procedure leads to corner plots as an output, of which an example is shown in Fig.~\ref{fig:cornerplot} for two pulsars. On the left panel, for pulsar J0740$+$6620 and on the right panel for J1514$-$4946. The error bars on the two characteristic angles are deduced from the quartiles at 99.85\% and 0.15\%, corresponding to a $3\,\sigma$ confidence interval. For J0740$+$6620 which is an example of a good fitting, the p-value passes the KS test. However, as a less accurate fitting example, we also show J1514$-$4946, on the right panel of Fig.~\ref{fig:cornerplot}.

\subsection{Pulsars with symmetrical solution kept}\label{appendix:first-fit}

\begin{figure*}[h]
    \centering
    \includegraphics[width=0.8\linewidth]{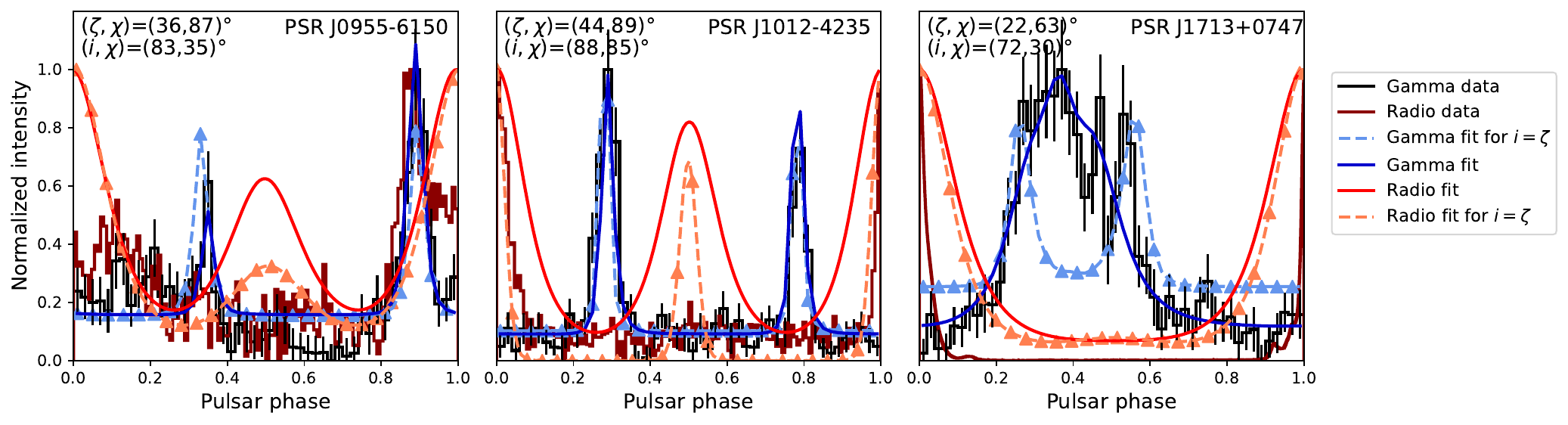}
    \caption{Original solution for pulsars from the first sample for which the symmetrical solution has been kept.}
    \label{fig:MSP1-firstfit}
\end{figure*}

\begin{figure*}[h]
    \centering
    \includegraphics[width=\linewidth]{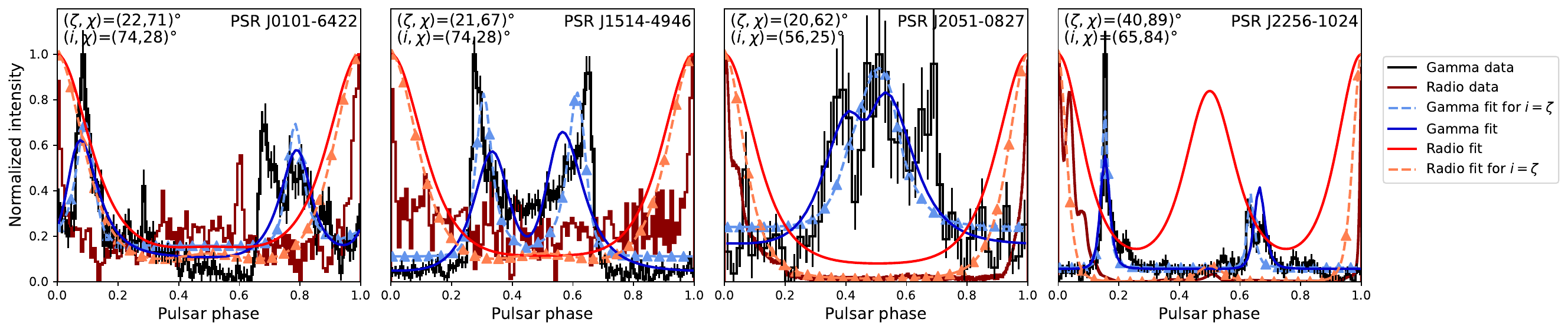}
    \caption{Original solution for pulsars from the second sample for which the symmetrical solution has been kept.}
    \label{fig:MSP2-firstfit}
\end{figure*}

To demonstrate the small impact of taking an almost symmetrical solution for the $\gamma$-ray fitting, we show here the light curves using the $\theta_{\rm best\ fit}$ that minimize $\rchi^2$ on the whole parameter space for PSR J0955$-$6150, J1012$-$4235, J1713$+$0747 from the first sample, see figure~\ref{fig:MSP1-firstfit}, and for PSR J0101$-$6422, J1514$-$4946, J2051$-$0827 and J2256$-$1024 from the second sample, see figure~\ref{fig:MSP2-firstfit}. The feature of these light curves are indeed very similar to the one obtained by swapping the angles $\rchi$ and $\zeta$.

\section{\cor{The cross-correlation method}\label{appendix:crosscorrelation}}

In this section we explain the principle of the normalized cross-correlation as delineated in \cite{lewis1995fast} and \cite{jacovitti1993normalized}. We substantiate this method through a concise mathematical proof, tailored specifically to our context.

\subsection{Principle}

Let us consider \( u \) and \( v \) as two discrete periodic signals of length~\( N \). We define \( v_t \) as the time-shifted signal \( v_{t}[k] = v[k-t] \), where the minus operation is performed modulo \( N \) due to the periodicity. Our objective is to apply three transformations to \( v \), namely: time-shift, spatial translation, and spatial scaling, such that it aligns as closely as possible with the signal \( u \). This objective is quantitatively accomplished by minimizing the following expression
\begin{equation}
    \|  u - (a \, v_t + b)\|^2= \sum_k \big(u[k] - a \, v[k-t] - b\big)^2
\end{equation} 
over all possible parameters $t\in \mathbb Z,a\in \mathbb R_+,b\in \mathbb R$.

\textbf{Proposition:}  Let us define the centred versions of our signals as $\dot u= u - \mathtt{mean}(u)$ and $\dot v= v - \mathtt{mean}(u)$. The minimizers of $\| u - (a \, v_t + b)\|^2$  are given by   $\hat t,\hat a,\hat b$ where $\hat t$ and  $\hat a$ are the location and the value of the maximum of the function
\begin{equation}\label{cross-corr}
t\to \frac{1}{K} \sum_k \dot u[k] \dot v[k-t]  \qquad \text{ with } K=\sum_k (\dot v[k])^2
\end{equation}
and where $\hat b= \mathtt{mean}(u) - \hat a \,  \mathtt{mean}(v)$.

Observation: The function \ref{cross-corr} is called the cross-correlation, and is calculated efficiently utilizing the Fast Fourier Transform (FFT) algorithm. The comprehensive algorithm is detailed in the concluding subsection.

\subsection{Proof}

To establish the proposition, we reduce to the case where the signals are centred, with the help of the following lemma.

\textbf{Lemma:}  The minimizers $\dot{\hat t}, \dot{\hat a}, \dot{\hat b}$ of  $\|  \dot u - a \, \dot v_t - b\|^2$ are linked to the minimizers $\hat t,\hat a,\hat b$  of $\|  u - a \, v_t - b\|^2$  by: 
\begin{equation}
    \dot{\hat t}=\hat t, \quad \dot{\hat a}=\hat a , \quad  \dot{\hat b} = \hat b - c
\end{equation}
with $c=\mathtt{mean}(u)-\hat a\,\mathtt{mean}(v)$. 

\noindent 
{\it Proof of the lemma:} Injecting $\dot{\hat t}, \dot{\hat a}, \dot{\hat b}$ in the expression they need to minimize, we get
\begin{equation}
     \|  \dot u - \dot{\hat a} \, \dot v_{\dot{\hat t}} - \dot{\hat b}\|^2=  \|  \dot u - \hat a \, \dot v_{\hat t} - \hat b + c\|^2
     = \|   u - \hat a \, v_{\hat t} - \hat b \|^2
\end{equation}
which is actually the minimal possible value from the very definition of the triplet $\hat t, \hat a, \hat b$. $\square$

Building upon this lemma, it becomes sufficient to demonstrate the proposition within the  context where $u$ and $v$ have already been centred. For the remainder of the proof, we will operate under this assumption, thus rendering the dot notations unnecessary.

\noindent 
{\it Proof of the proposition:} Initially, we address the minimization problem for a specified value of \( t \). Thus, we seek
$$ \label{least_square}
\hat a_t, \hat b_t = \underset{a,b}{\text{argmin}}  \|  u - a  \, v_t - b\|^2  \ .
$$
This is a classical least squares problem that is resolved utilizing the normal equation. The details are as follows: Let \( V_t \) represent the matrix wherein the first column consists of coefficients equal to 1, and the second column is the vector \( v_t \). Let \( U \) denote the matrix comprising a single column, which is the vector \( u \). According to the normal equation, the solution to \ref{least_square} is provided by
 $$
\begin{pmatrix} \hat b_t \\ \hat a_t \end{pmatrix}= (V^T_t \, V_t)^{-1} \, V_t^T \, U \ .
$$
Given that both $u$ and $v$ have been centred, the calculation process is straightforward
$$
\begin{pmatrix} \hat b_t \\ \hat a_t \end{pmatrix}= \begin{pmatrix} 0 \\
 \frac{1}{K} \sum_k u[k] \, v[k-t] \end{pmatrix} \ .
$$
Now, to find the $t$-minimizer, we have to minimize ${t \to  \|  u - \hat a_t v_t \|^2}$. Let's develop
\begin{align*}
\hat t& = \underset{t}{\text{argmin}}\sum_k\Big(   u[k] - \hat a_t \, v_t[k]  \Big)^2   \\
&= \underset{t}{\text{argmin}}\Big[ \sum_k  (u[k])^2  + (\hat a_t)^2 (v[k-t])^2 - 2 \, u[k] \, \hat a_t \, v[k-t] \Big]\\
&= \underset{t}{\text{argmin}} \Big[(\hat a_t)^2  \sum_k (v[k-t])^2 - 2 \, \hat a_t \sum_k u[k] \, v[k-t]  \Big]\\
&= \underset{t}{\text{argmin}} \Big[(\hat a_t)^2 K - 2 \hat a_t \, K \, \hat a_t\Big]\\
&= \underset{t}{\text{argmax}}\  (\hat a_t)^2 = \underset{t}{\text{argmax}}\  \hat a_t
\end{align*}
which concludes the proof of the proposition. $\square$ 

\subsection{Algorithm}
Analogously to mathematical reasoning, the algorithm reduces to the case where vectors $u$ and $v$ are centred. For concreteness, we give Python code to perform this cross-correlation.
\begin{lstlisting}[caption={python import}, label={lst:import}]
from numpy.fft import fft,ifft
from numpy import mean,sum,roll,flip,argmax,real
\end{lstlisting}

\begin{lstlisting}[caption={Preliminary function: which suppose that inputs u and v are centred}, label={lst:centred}]
def find_best_shift_scaling_for_centred(u, v):
    #v_inverse[i]=v[-i] (with periodicity)
    v_inverse = roll(flip(v), 1) 
    K = sum(v ** 2)
    cross_correlation = real(ifft(fft(u) * fft(v_inverse))) / K
    t = argmax(cross_correlation)
    a = cross_correlation[t]
    return t, a
\end{lstlisting}

\begin{lstlisting}[caption={Final function using the previous one. }, label={lst:full}]
def find_best_shift_scaling_translation(u, v):
    mean_u = mean(u)
    mean_v = mean(v)
    t, a = find_best_shift_scaling_for_centred(u - mean_u, v - mean_v)
    b = mean_u - a * mean_v
    return t, a, b
\end{lstlisting}

\begin{figure}[h]
    \includegraphics[width=\linewidth]{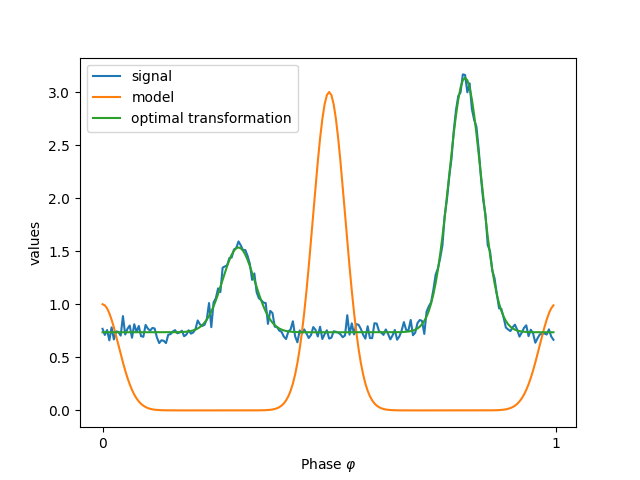}
    \caption{A graphical evaluation of the proposed algorithm. The orange curve represents a periodic signal modeling a reference signal from our atlas. The blue curve denotes a noisy signal corresponding to an observed measurement. The green curve illustrates the optimal transformation of the orange signal —obtained through temporal shifts, spatial shift and amplitude scaling— that best matches the blue signal.
    }
    \label{fig:placeholder}
\end{figure}

Fig.~\ref{fig:placeholder} shows an example of fitting to a periodic and double peaked signal. To draw this plot, we compute $u$ by taking $\varphi$, a vector of 200~phases from $0$ to $2\pi$. Then, we derive $x=\cos(\varphi), y=\sin(\varphi)$. Finally, we set $u=g(x-1,y) + 3 g(x+1,y)$ where $g(x,y) = \exp(-(x^2+y^2)/\sigma^2)$ is a Gaussian curve with a small standard deviation $\sigma^2=0.1$. 



\end{document}